%% file: Gms.tex
\documentclass[nospthms, smallextended, final]{svjour2}
\usepackage[utf8]{inputenc}
\usepackage{mathptmx}

\usepackage{color}
\usepackage{graphicx}
\graphicspath{{Figs/}}
\usepackage{amsmath}
\usepackage{amsthm}
\usepackage{amssymb}
\usepackage{braket}
\usepackage{cancel}

\input{notations4}

\begin{document}
\title{General limit distributions for sums of random variables with a matrix product representation}
\date{\today}
\author{Florian Angeletti \and Eric Bertin \and Patrice Abry}
\institute{ Florian Angeletti \at National Institute for Theoretical Physics (NITheP), Stellenbosch 7600, South Africa 
\at Institute of Theoretical Physics, University of Stellenbosch, Stellenbosch 7600, South Africa 
\\\email{florianangeletti@sun.ac.za}
\and Eric Bertin \at Laboratoire Interdisciplinaire de Physique, Universit\'e Joseph Fourier Grenoble, CNRS UMR 5588, BP 87, F-38402 Saint-Martin d'H\`eres, France 
\\\email{eric.bertin@ujf-grenoble.fr}
\and Patrice Abry \at Universit\'e de Lyon, Laboratoire de Physique, ENS Lyon, CNRS,
46 All\'ee d'Italie, F-69007 Lyon, France 
\\\email{patrice.abry@ens-lyon.fr}
}
\maketitle
\begin{abstract}
The general limit distributions of the sum of random variables described by a finite matrix product ansatz are characterized.
Using a mapping to a Hidden Markov Chain formalism, non-standard limit distributions are obtained, and related to a form of ergodicity breaking 
in the underlying non-homogeneous Hidden Markov Chain.
The link between ergodicity and limit distributions is detailed and used
to provide a full algorithmic characterization of the general limit distributions.   
\end{abstract}

\section{Introduction}

The statistics of sums of random variables play a major role in statistical physics, through the law of large numbers and the central limit theorem, which justify the existence of deterministic values of macroscopic observables in large systems, as well as the Gaussian statistics of the very small fluctuations around these deterministic values.
Probabilistic notions like random walks, which are intimately related to the central limit theorem \cite{FellerI-66}, have also found applications in many different fields.
Beyond standard formulations of the law of large numbers and of the central limit theorem \cite{Gnedenko54,FellerI-66,FellerII-66}, valid for independent and identically distributed ($\iid$) random variables, different types of generalizations including correlations between random variables, or considering non-identically distributed variables, have been proposed \cite{FellerII-66,Petrov95}.
At a qualitative level, these theorems are expected to be valid as long as correlations are not too strong, and as long as the statistics of individual variables does not differ too much one from the other.
When considering nonequilibrium systems, such assumptions may however not be valid.
For instance, the relevant physical observable may have an infinite mean value, as in the case of laser cooling \cite{Bardou02} or aging \cite{Bouchaud90,Bouchaud92} phenomena, leading to a breakdown of the law of large numbers. A similar breakdown occurs in the presence of long-range correlations, as seen in active systems \cite{Ramaswamy03,Toner05} and in boundary driven systems \cite{DerridaEvans93,Essler96}.
Broad distributions and long-range correlations also lead to a breakdown of the standard central limit theorem. If the variables have infinite variance, the Gaussian limit distribution is replaced by L\'evy stable laws \cite{Gnedenko54}, which have found many applications in connection to anomalous diffusion processes \cite{Bouchaud90,Metzler00}.
If the random variables have strongly different statistics while still being independent, as in the case of the $1/f^{\alpha }$-noise problem \cite{Antal02} or of the XY-model at low temperature \cite{Bramwell01} in Fourier space, non-Gaussian distributions also appear \cite{Clusel08}.
The most difficult case is probably that of strongly correlated variables, since the independence property allows for important simplifications in the calculation of the distribution of the sum.
An important example stems from transforms of long-range dependent gaussian processes, 
for which generalizations of the central limit theorem have been obtained \cite{Taqqu79,Rosenblatt81}. 
In the study of statistical physics models, especially when considering exact solutions, another class of random variables however plays an increasingly important role, namely correlated random variables described by a matrix product ansatz.
Such stochastic models have now become a standard tool to describe the exact steady-state probability distributions of one-dimensional nonequilibrium models, ranging from the Asymmetric Simple Exclusion Process (ASEP) and its generalizations \cite{Hakim83,DerridaASEP93,Essler96,Mallick97,Evans07,Mallick09,Crampe2011,Lazarescu2011,Lazarescu2013} to different kinds of reaction-diffusion processes \cite{Hinrichsen96,Hinrichsen00,Jafarpour03,Jafarpour04,Hieida04,Evans07,Basu09,Hinrichsen13} and to KPZ equations \cite{Spohn13}.
The use of infinite matrices is often required, for instance in the case of the ASEP model, but a significant number of models can however be solved with finite matrices. This is the case of reaction-diffusion models \cite{Jafarpour03,Jafarpour04,Hinrichsen96,Hinrichsen00,Hieida04,Evans07,Basu09,Hinrichsen13}, coupled KPZ equations \cite{Spohn13}, or even the ASEP model if some constraints between parameters are imposed \cite{Essler96,Speer97,Mallick97}.

In this contribution, we aim at determining the limit distributions of the sum of random variables described by a matrix product ansatz with finite matrices, either for discrete \cite{DerridaASEP93,Evans07} or continuous \cite{Angeletti2012:ICASSP,Angeletti2013:PiMat} variables.
This contribution complements an earlier publication \cite{Angeletti2013:MatSum:Letter} dedicated to specific, yet representative cases, and provides general results for all types of finite matrices.
In section \ref{sec:presentation}, we define a random vector with matrix representation 
and underline the advantages of the Hidden Markov Chain representation for the study of the statistics of the sum. General expressions of the limit distributions corresponding to the law of large numbers and of the central limit theorem, involving an auxiliary distribution $Q$ characterizing the Hidden Markov Chain, are derived.
In Sect.~\ref{sec:methodo}, the general methodology is presented.
Sections \ref{sec:tir} and \ref{sec:reduced} jointly establish the main results of this article.
First, section~\ref{sec:tir} analyzes a subclass of matrix representation models referred to as ``totally irreversible models'', for which the distribution $Q$ can be determined explicitly.
Limit distributions of the sum are thoroughly determined for this subclass. Then, section~\ref{sec:reduced} establishes that the sum of any matrix-correlated random vector is statistically equivalent to
the sum of a ``reduced'' model, that falls into the class of totally irreversible models. A precise mapping to the reduced model is given, thus providing explicit generalizations of the law of large numbers and of the central limit theorem for an arbitrary random vector described by a matrix product ansatz (with finite matrices).
In order to facilitate practical applications, we illustrate with a concrete example the steps needed to compute the limit distributions
in section~\ref{sec:algo}.

\section{Random variables described by a matrix product ansatz} \label{sec:presentation}
\subsection{Matrix product representation}
We study the sum
\begin{equation}
S(\vec X)= \sum_{i=1}^{N} X_i 
\end{equation}
of random variables $\vec X=(X_1,\dots,X_N)$ whose joint probability
density function is described by a matrix product ansatz, namely\footnote{The definition~\eqref{eq:Def} is valid for any probability space, however in the present article, we restrict ourselves to real random variables.}
\begin{equation}
\label{eq:Def}
P(x_1,\dots,x_N) = \frac{1}{\Lf (\mE^N)}\, \Lf \p{ \mR(x_1) \mR(x_2) \dots \mR(x_N) }
\end{equation}
where $\mR(x)$ is a $D \times D$ matrix function with real
nonnegative entries, the matrix $\mE$ is defined as
$\mE=\int_{-\infty }^{\infty } \mR(x)\,dx$,
and $\Lf$ is a linear form defined as
\begin{equation}
\label{eq:def-L}
\Lf\p{M} = \tr\p{\mA^T M},
\end{equation}
with $\mA$ a given $D \times  D$, nonzero matrix with real nonnegative
entries. We further assume that for all $N \ge 1$, $\Lf(\mE^N) \ne  0$.
This ansatz, first proposed in \cite{Angeletti2012:ICASSP,Angeletti2013:PiMat}, 
is a generalization of the standard forms used in statistical physics \cite{Evans07}. These
standard forms can be recovered by choosing $\mA$ as the identity matrix, or as
$\mA = V W^T$ so that $\Lf(M) = \bra V M  \ket W$.
Eq.~\eqref{eq:Def} is a natural generalization
to correlated variables of the i.i.d.~case, replacing the product of
real functions by a product of matrix functions.

As shown in \cite{Angeletti2012:ICASSP,Angeletti2013:PiMat}, the joint probability
(\ref{eq:Def}) can be reinterpreted within the framework of Hidden Markov Chains
\cite{Cappe05}. To this aim, we introduce a Markov chain $\Gamma  \in  \Set{1,\dots, D}^{N+1}$ such
that
\begin{align}
\label{eqn:trans0}
\Prob (\Gamma_1 = i,\Gamma_{N+1} = f) &= \mA_{if} \frac{(\mE^N)_{if}}{\Lf(\mE^N)} \; , \\
 \label{eq:transition}
\Prob (\Gamma_{k+1}= j | \Gamma_{k}=i, \, \Gamma_{N+1} = f ) &= \mE_{ij} \frac{(\mE^{N-k})_{jf}}{(\mE^{N-k+1})_{if}} \; , \quad k=1,\dots,N.
\end{align}
Note that this Markov chain is non-homogeneous and of a nonstandard type, due to the dependence on the final state $\Gamma_{N+1}$.
In particular for $k=N$, the transition rate  $\Prob (\Gamma_{k+1}= j | \Gamma_{k}=i, \, \Gamma_{N+1} = f )$
equals $1$ if $j=f$ and $0$ otherwise.
Combining Eqs.~\eqref{eqn:trans0} and \eqref{eq:transition}, the global probability of a given chain $\Gamma $ reads
\begin{equation} \label{eq:kappa}
\Prob \p{ \Gamma  } = \frac{\mA_{\Gamma_1 \Gamma_{N+1}}}{\Lf\p{\mE^N}}\,
\mE_{\Gamma_1 \Gamma_2} \mE_{\Gamma_2 \Gamma_3} \dots \mE_{\Gamma_{N-1} \Gamma_N} \mE_{\Gamma_N \Gamma_{N+1}} \; .
\end{equation}
For a given $\Gamma$, the random variables $(X_1,\dots,X_N)$ are independent but non-identically distributed, 
with a probability distribution depending on $\Gamma $:
\begin{equation}\label{eq:X|gamma}
\Prob (x_1,\dots,x_N | \Gamma  ) = \prod_{k=1}^N \mP_{\Gamma_k \Gamma_{k+1}}(x_k)
\end{equation}
where $\mP_{ij}(x)$ is a probability density defined, for all $(i,j)$, as\footnote{Note that $\mP_{ij}(x)$ is uniquely defined only when $\mE_{ij} \ne 0$. When $\mE_{ij} = 0$, the distribution $\mP_{ij}(x)$ plays no role and can thus be any arbitrary distribution.}
\begin{equation}
 \mE_{ij} \mP_{ij}(x) = \mR_{ij} \p{x} \; .
\end{equation}
As a result, the distribution $\Prob (x_1,\dots,x_N)$ can be written as a mixture of factorized distributions,
\begin{equation}
\Prob (x_1,\dots,x_N) = \sum_{\Gamma} \Prob \p{ \Gamma }\, \Prob (x_1,\dots,x_N | \Gamma)\, .
\end{equation}
This formulation using a hidden Markov chain $\Gamma $ is equivalent
to the definition Eq.~\eqref{eq:Def} using matrices \cite{Angeletti2012:ICASSP,Angeletti2013:PiMat}.
This yields a procedure to simulate the correlated random variables described
by Eq.~\eqref{eq:Def} \cite{Angeletti2013:PiMat}: (i) $\Gamma_1$ and $\Gamma_{N+1}$
are chosen at random according to distribution \eqref{eqn:trans0};
(ii) the random chain $\Gamma $ is obtained from transition rates
\eqref{eq:transition}; (iii) the random variables $X_k$, $k=1,\dots N$,
are drawn randomly from the distributions $\mP_{\Gamma_k \Gamma_{k+1}}(x_k)$.

As seen in Eq.~\eqref{eq:X|gamma}, for a fixed $\Gamma $, the random variables
$(X_1,\dots,X_N)$ are independent. Correlations, when present, thus emerge from
the correlations within the hidden chain $\Gamma $ and the mixture of distributions $\Prob (x_1,\dots,x_N | \Gamma)$.

\subsection{Statistics of the sum}

This separation of randomness between the independent random vector $(\vec X|\Gamma )$ and the hidden Markov chain
$\Gamma $ plays a key role in our analysis of $S(\vec X)$. Specifically, the distribution of $S(\vec X)$ can be determined by
first computing the distribution of $S( \vec X | \Gamma ) =\sum_{k=1}^N X_k$, where
the $X_k$'s are drawn from $\Prob (x_1,\dots,x_N | \Gamma )$, and then averaging
the distribution of $S(\vec X | \Gamma )$ over $\Gamma $.

For a given chain $\Gamma $, we introduce the transition frequencies $\nu_{ij}$
from $i$ to $j$ in $\Gamma $,
\begin{equation}
\nu_{ij} = \frac{1}{N}\, \card \Set{k | \Gamma_k=i \; \text{and} \;
\Gamma_{k+1}=j }\,.
\end{equation}
The sum $S(\vec X | \Gamma  ) $ can be rewritten as
\begin{equation} \label{eq:sum:rew}
S( \vec X | \Gamma  ) = \sum_{i,j=1}^D \sum_{k=1}^{N \nu_{ij}} \cX{ij}_k
= S(\vec X |\nu)
\end{equation}
with $\nu = (\nu_{11},\nu_{12}\dots,\nu_{DD})$, and where the variables $\cX{ij}_{k}$, $k=1,\dots,N\nu_{ij}$, are i.i.d.~random
variables drawn from the distribution $\mP_{ij}(x)$.
The statistics of the sum therefore only depends on the transition frequencies $\nu_{ij}$. We are specifically interested in the validity
of the law of large numbers and of the central limit theorem.
If we conjecture that the transition frequencies $\nu $ admit a limit distribution $Q(\nu )$,
an analogue of the law of large numbers can be derived \cite{Angeletti2013:MatSum:Letter}
\begin{equation} \label{eq:Psi-s}
\Psi (s) \equiv  p \p{ \frac{S(\vec X)} N = s } =  \int  Q(\nu )  \delta  \p{ s - \sum_{i,j=1}^D \nu_{ij} \mu_{ij}  } \prod_{i,j=1}^D d\nu_{ij}  
\end{equation}
where $\mu_{ij} = \langle\cX{ij}\rangle$.
The sample average $S(X)/N$ therefore converges to a mixture of Dirac distributions.  

The law of large numbers holds either when all $\mu_{ij}$'s are equal
(or at least those associated to nonzero $\nu_{ij}$), or
when the empirical frequencies $\nu_{ij}$ converge to nonrandom values
$\bar\nu_{ij}$ in the limit $N\rightarrow  \infty $, in which case
the rescaled sum $s$ converges to the deterministic limit
$\mu =\sum_{i,j} \pnu{ij} \mu_{ij}$.
When the law of large numbers is satisfied, an analogue of the
central limit theorem can be derived
\begin{equation} \label{eq:Phi-z}
\Phi (z) = p\p{ \frac{S(\vec X) - N \mu } { \sqrt N }  = z } =  \int  \frac{\prod_{i,j} d\nu_{ij}\, Q(\nu )}{\sqrt {2\pi  \sum_{i,j} \nu_{ij} \sigma_{ij}^2}} \, e^{-z^2/(2\sum_{i,j} \nu_{ij} \sigma_{ij}^2)} \; ,
\end{equation}
assuming that all variables $\cX{ij}$ have a finite variance $\sigma_{ij}^2$, given by
\begin{equation}
\sigma_{ij}^2 = \Esp{\p{\cX{i,j}}^2} -  \Esp{\cX{i,j}}^2.
\end{equation}  
The limit distribution $\Phi (z)$ of the rescaled sum $z$ can thus be expressed as a mixture of Gaussian distributions of variance $\sum_{i,j} \nu_{ij} \sigma_{ij}^2$, each term in the mixture corresponding to different values of $\nu_{ij}$'s. As a result, the central limit theorem is valid when the variance $\sum_{i,j} \nu_{ij} \sigma_{ij}^2$ takes the same value for all sequences $\nu$ of frequencies having a nonzero probability $Q(\nu )$.
This happens either when all the individual variances $\sigma_{ij}$,
associated to a nonzero value $\nu_{ij}$, are equal or when $\nu$
takes a non-random value. In this latter case, the distribution
$Q(\nu)$ is a Dirac distribution around a particular sequence of frequencies $\bar\nu$,
and there is a single term in the mixture.

Accordingly, the distribution $Q(\nu)$ turns out to be a key element
to characterize the limit distributions $\Psi (s)$ and, when the law of large numbers holds, $\Phi (z)$. When the distribution $Q(\nu)$ is known, the limit distributions $\Psi (s)$ and $\Phi (z)$ can be obtained from \Eqref{eq:Psi-s} and \eqref{eq:Phi-z} respectively.
One of the difficulties in the determination of $Q(\nu)$ resides in the non-homogeneous nature of the chain $\Gamma $ in the general case.
In the following sections, we provide a general framework to determine the distribution $Q(\nu)$. The general methodology is presented in Sect.~\ref{sec:methodo}. Sect.~\ref{sec:tir} focuses on a specific, yet important class of matrices $\mE$ that we denote as ``totally irreversible models'', for which the distribution $Q(\nu)$ can be determined in a relatively straightforward way. Then Sect.~\ref{sec:reduced} shows how the general case can be mapped to a reduced model belonging to the class of ``totally irreversible models''.

\section{General methodology}
\label{sec:methodo}
\subsection{Generic form of the matrix $\mE$}

In order to understand the non-homogeneous behavior of $\Gamma$ and to determine the distribution $Q(\nu)$, it is useful
to note that the shape of the matrix $\mE$ imposes global constraints on the 
hidden Markov chain $\Gamma $.

More specifically, one of the consequences of the Perron-Frobenius theorem \cite{Seneta06} is that any non-negative matrix $\mE$ can be decomposed into a block upper triangular matrix using only a relabelling of indices (see section~\ref{sec:algo}) 
\begin{equation} \label{eq:Perron}
\mE=
\begin{pmatrix}
\mB{1} &    *   & *      & * \\  
0      & \ddots & *      & *        \\
\vdots & \ddots & \ddots & *        \\
0      & \cdots &    0   & \mB{\nc}   \\ 
\end{pmatrix}.
\end{equation}
The matrices $\mB{c}$, $c=1,\dots,p$, are irreducible square matrices of size $D_c$, with $ \sum_{c=1}^{\nc} D_c = D$.
The irreducibility of the block $\mB{c}$ can be characterized as
\begin{equation} \label{eq:def:irr}
\forall  (i,j) \in  \Set{1,\dots, D_c}^2, \quad \exists m\in\N,\, \p{\mB{c}^m}_{i,j} >0.
\end{equation}
Blocks represented by the symbol $*$ are arbitrary at this stage.
The $\nc$ irreducible blocks $\mB{c}$ partition the indices of the matrix $\mE$ into $\nc$ subsets, or classes $\sB{c}$, $c=1,\dots,\nc$, such that $\sB{c}$ is the set of indices of the block $\mB{c}$ in the matrix $\mE$.
Formally, $\sB{c}$ reads
\begin{equation}
\sB{c} 
= \Set{1+\sum_{c'=1}^{c-1} D_{c'}, \dots, \sum_{c'=1}^{c} D_{c'} } \; .
\end{equation}

This decomposition of $\mE$ has two advantages. First, 
the spectrum $\lambda_1, \dots, \lambda_m$ of $\mE$ is the union of the spectra $\{ \lambda_{c,l} \}$ of the blocks $\mB{c}$,
\begin{equation}
\Set{ \lambda_1, \dots, \lambda_m } =  \bigcup_{c,l}  \Set{ \lambda_{c,l} } \; .
\end{equation}  
Moreover, the Perron-Frobenius theorem states that any irreducible matrix $\mB{c}$ admits a dominant real positive eigenvalue $\Lambda_c$ such that 
for any $l$
\begin{equation} 
\Lambda_c \ge |\lambda_{c,l} | \;.
\end{equation}
The dominant eigenvalue $\Lambda $ of $\mE$ is among these block dominant eigenvalues $\Lambda_c$
\begin{equation} \label{eq:max:Lambda}
\Lambda  = \max_{c=1,\dots,p} \{ \Lambda_c \} \;.
\end{equation}

\subsection{Connectivity of $\mE$ and global structure of $\Gamma $}

The Perron-Frobenius decomposition of $\mE$ can be interpreted as a rough description of the connectivity of the matrix $\mE$. Consider an oriented
graph $\gE(\mE)$ whose vertices are the values $i=1,\dots,D$, and where two vertices $(i,j)$ are connected by an edge 
if and only if $\mE_{ij} \ne  0$. Combined with \Eqref{eq:transition}, this implies that if the transition $i \rightarrow  j$ in the chain $\Gamma $ has a non-zero probability, then there is an edge between $i$ and $j$.
Going a little further, an equivalent characterization of the irreducibility of the block $\mB{c}$ is that
for any couple of vertices $(i,j) \in  \sB{c}$, there is a sequence of edges $i \rightarrow  \dots \rightarrow  j$. 
The classes $\sB{c}$ partition the indices of the matrix $\mE$ into subsets inside which every transition is reversible, in the sense that if there exists a sequence of edges from $i$ to $j$, another sequence of edges from $j$ to $i$ also exists.
This is a kind of ergodic property for the chain $\Gamma $. By contrast, if we know that the chain $\Gamma $ goes from $i \in  \sB{c}$ to $j \in  \sB{c'}$
with $c\ne  c'$ then the structure of $\mE$ implies that, necessarily, $c < c'$.
This transition is irreversible and the ergodicity of  the chain $\Gamma $ is then broken. 

Taking advantage of the reversibility of transitions within blocks, we will use in the following a two-level description of the chain $\Gamma$, by distinguishing reversible transitions inside blocks, and irreversible transitions between different blocks. Let us first note that any chain $\Gamma $ with a non-zero probability can be written as
\begin{equation} \label{eq:chain:gener}
\Gamma   = (i_{c_1,1}, \dots, i_{c_1,n_1},
\cdots, i_{c_r,1}, \dots, i_{c_r,n_r}) \; , 
\end{equation}
where the indices $i_{c,k}$, $k=1,\dots,n_c$, belong to $\sB{c}$.
The number of distinct classes 'visited' by $\Gamma$ satisfies $1\le r\le \nc$.
We can then coarse-grain the chain $\Gamma$ by replacing, for all $k$, $\Gamma_k$ by the index $\pfC_k$ of the class of indices to which $\Gamma_k$ belongs
(formally, $\pfC_k = c \iff \Gamma_k \in  \sB{c}$).
We call 'class chain' the resulting coarse-grained chain $\pfC$, which reads,
from Eq.~(\ref{eq:chain:gener}),
\begin{equation} \label{eq:cshape}
\pfC  = (\repeated{c_1}{n_1}, \dots, \repeated{c_r}{n_r}), \qquad  c_1 < c_2 < \dots < c_r.     
\end{equation}
Looking at Eq.~(\ref{eq:cshape}), a natural step to further coarse-grain the class chain $\pfC$ is to keep only the list of distinct classes within $\pfC$ (thus loosing information on the 'time' spend by $\Gamma$ within each class).
We denote as 'structure chain' the chain $\pC$ of distinct class indices,
\begin{equation} \label{eq:cstructure}
\pC  = (c_1, \dots, c_r), \qquad  c_1 < c_2 < \dots < c_r.     
\end{equation}
This notion of structure chain will be useful in the following.
Note that the length of the chain $\pC$ is not fixed, and will be denoted $\len{\pC}$ to emphasize its dependence on $\pC$ 
in the following.
Similarly, using \Eqref{eqn:trans0}, the non-zero entries of the matrix $\mA$ can be interpreted
as the admissible pairs of initial and final states for the chain $\Gamma$ in the graph $G(\mE)$. This gives us supplementary constraints
on the shape of $\Gamma$. In particular, we call reachable a class $\sC{c}$ for which there is a path 
from $i$ to $f$ with $\mA_{if}>0$ passing through $\sC{c}$:
\begin{equation}
\exists k \in \sC{c}, \exists (i,f) \in \Set{  A_{if} >0 }, \quad i \rightarrow \dotsm \rightarrow k \rightarrow \dotsm \rightarrow f  
\end{equation}
If a class $\sC{c}$ is not reachable, there is no chain $\Gamma$ with a non-zero probability passing through this class.
It is thefore possible to remove the rows and columns of indices $\sC{c}$
from the matrices $\mA$ and $\mE$ without altering the joint probability density function defined in \Eqref{eq:Def}. Intuitively, unreachable
classes correspond to unused parts of the matrices $\mE$ due to restrictions on the pairs of initial and final states imposed by $\mA$. Without loss of generality, we consider in the following only pairs of matrices $(\mA,\mE)$ with no unreachable classes. 

To sum up, the transitions within the chain $\Gamma $ can therefore be divided into two groups:
reversible transitions inside a block $\mB{c}$ and irreversible transitions 
between the classes $\Xi_c$. If there are no irreversible transitions,
$\mE$ is irreducible.
In this situation, $\S(\vec X)$ converges towards a classical limit distribution \cite{Angeletti2013:MatSum:Letter}.
Irreversible transitions must therefore play a major role in the emergence of non-standard distributions.

\subsection{Principle of the determination of $Q(\nu)$}

The method we use to determine the distribution of frequencies $Q(\nu)$ can be summarized as follows. The dynamics of the chain $\Gamma$ within the irreducible blocks $\mB{c}$ is known to be ergodic \cite{Angeletti2013:MatSum:Letter}. A natural idea is thus to define a coarse-grained dynamics of $\Gamma$ in terms of the class chain $\pfC$ (and its associated structure chain $\pC$) defined in \Eqref{eq:cshape}, replacing the internal dynamics within blocks by a simpler, effective dynamics. This is the topic of Sect.~\ref{sec:reduced}.

The dynamics of the class chain $\pfC$ can be shown to correspond to a subset of the possible dynamics of $\Gamma$, described by a subclass of matrices $\mE$ that we call ``totally irreversible models'' (see Sect.~\ref{sec:tir}).
Thus the generic case of random vectors described by a matrix-product ansatz \eqref{eq:Def} can be mapped onto the subclass of totally irreversible models.

This latter class can be characterized thoroughly in a relatively simple way, considering the limit of an infinite number of random variables (or equivalently, an infinite length of the chain $\Gamma$).
The determination of the distribution $Q(\nu)$ for totally irreversible models can be done in two steps. First, the conditional distribution $Q(\nu|\pC)$, restricted to a given structure chain $\pC$, can be obtained as a flat measure over the values of $\nu$ allowed by 'geometrical' constraints --see \Eqref{eq:Qnu-C}.
Second, the full distribution $Q(\nu)$ is obtained as an average of $Q(\nu|\pC)$ over all (maximal length) structure chains $\pC$, as described by Eqs.~\eqref{eq:Qnu-av} and \eqref{eq:FDlim} below.

In the following, we first characterize in details the class of totally irreversible models (Sect.~\ref{sec:tir}) and then show how the general case can be mapped to this specific class (Sect.~\ref{sec:reduced}).

\section{Totally irreversible models} \label{sec:tir}

We can use the above dual nature of the transitions of the chain $\Gamma$ to study separately the effect of
the inner structure of the block $\mB{c}$, and of the transitions between these blocks (or in other words, between the classes $\Xi_c$) on the limit distributions.

\subsection{Definition and properties}

As a first step, let us consider the subclass of 'totally irreversible' matrices $\mE$ for which
the irreducible classes $\Xi_c$ reduce to a single element, the singletons $\Set c$.
In this case, the inner structure of the block $\mB{c}$ is trivial and the only reversible transitions are transitions of the form $ i \rightarrow i$.
By studying this class of totally irreversible matrices $\mE$, we can focus on the effect of irreversible transitions on the sum $S(\vec X)$.
The general structure of a totally irreversible matrix $\mE$ can be obtained by simplifying \Eqref{eq:Perron} into
\begin{equation} \label{eq:Perron:Simple}
\mE=
\begin{pmatrix}
1      &    *   & *      & * \\  
0      & \ddots & *      & *        \\
\vdots & \ddots & \ddots & *        \\
0      & \cdots &    0   & 1   \\ 
\end{pmatrix}.
\end{equation}
Here, we have assumed that all diagonal elements are equal, and can thus be set to $1$ by a simple rescaling. As we will show below in Sect.~\ref{sec:reduced}, more general situations can be recast into this form, as far as the statistics of the sum is concerned.
With the matrix $\mE$ given in Eq.~(\ref{eq:Perron:Simple}), the chain $\Gamma$ is identical to the associated chain of classes $\pfC$, since each class contains a single element,
so that the chains $\Gamma$ with non-zero probability take the form given in \Eqref{eq:cshape}.
Combining Eqs.~\eqref{eq:cshape} and~\eqref{eq:kappa} shows that the probability of $\Gamma $ depends only on its associated structure chain $\pC $ defined in \Eqref{eq:cstructure}, given that diagonal coefficients of the matrix $\mE$ are equal to $1$,
\begin{equation} \label{eq:pGamma-irrev}
\Prob(\Gamma) = \frac{\mA_{\pC_1,\pC_{\len{\pC}}}\prod_{k=1}^{\len{\pC}-1} \mE_{\pC_k,\pC_{k+1}}}{\Lf{(\mE^N)}} \;.
\end{equation} 
Consequently, all chains $\Gamma $ with the same structure chain $\pC $ are equiprobable.
In the limit $N \rightarrow \infty$, the transition frequencies $\vec{\nu }$ for a given $\pC $ therefore follow a uniform distribution on the manifold $\Sp_\pC $ defined by
\begin{equation} \label{def:manifold-MC}
 \Sp_\pC  \equiv   \Set{\nu  | \sum_{k=1}^{\len{\pC }} \nu_{\pC_k,\pC_k} = 1; \;
\forall  i \ne  j,\, \nu_{ij} =0; \; \forall  i \notin  \pC ,\, \nu_{ii} = 0; \;
\forall i, \, 0\le \nu_{ii} \le 1 }. 
\end{equation}
Since the definition of this manifold contains only linear constraints, it can be interpreted as an intersection of half-spaces and hyperplanes in
$\mathbb{R}^D$, also called a polytope.
One can note that, for $\len{\pC }= 2$, $3$ and $4$, $\Sp_\pC $ is respectively a segment, a triangle and 
a tetrahedron. More generally, for any $\len{\pC }$, $\Sp_{\pC }$ is a generalized $\p{\len{\pC }-1}$-dimensional triangle called a $\p{\len{\pC }-1}$-simplex.
If we call $\lU_{\pC }$ the uniform probability density on  $\Sp_{\pC }$ then  
\begin{equation} \label{eq:Qnu-C}
Q( \nu|\pC) = \lU_{\pC }(\nu ) \; .
\end{equation}
Note that by definition, $\lU_{\pC }(\nu)=0$ if $\nu \notin \Sp_\pC$.
The full distribution $Q(\nu)$ is then obtained as an average over all possible chains $\pC$, weighted by their associated probability $p(\pC)$,
\begin{equation} \label{eq:Qnu-av}
Q(\nu) = \sum_{\pC} p(\pC)\, Q(\nu|\pC) \;.
\end{equation}
From \Eqref{eq:pGamma-irrev}, the probability $p(\pC)$ of observing a given structure chain $\pC$ is obtained by summing $\Prob(\Gamma)$ over all chains $\Gamma$ associated to a given $\pC$, yielding
\begin{equation}
p(\pC )  = \frac{\mA_{\pC_1,\pC_{\len{\pC}}} \prod_{k=1}^{\len{\pC}-1} \mE_{\pC_k,\pC_{k+1}} }{\Lf{(\mE^N)}}  \binom{N}{ \len{\pC }-1 }  \;.
\end{equation}
In the limit $N\rightarrow  \infty$, one has
\begin{equation}
 \binom{N}{\len{\pC }-1} \sim \frac{N^{\len{\pC}-1}}{(\len{\pC}-1) !} \; .
\end{equation}
The structure chains of maximal length are therefore favored due to an
entropic effect, so that only chains of length $\lmax$ have to be retained in the limit distribution of $\nu$. One can then replace $N^{\len{\pC}-1}/(\len{\pC}-1)!$ by $N^{\lmax-1}/(\lmax-1)!$, and the distribution $Q(\nu)$ given in \Eqref{eq:Qnu-av} can be rewritten as
(again for $N\rightarrow \infty$)
\begin{equation} \label{eq:FDlim}
 Q(\nu ) = \sum_{\pC , \len{\pC} = \lmax}  p(\pC|\lmax) \lU_{\pC }(\nu )
\end{equation}
where the distribution of the chain $\pC$ conditioned to an arbitrary length $l$ is given by
\begin{equation} \label{eq:pClmax}
 p(\pC|l) = \frac{\mA_{\pC_1,\pC_l}\prod_{k=1}^{l-1} \mE_{\pC_k,\pC_{k+1}}}{\Lf{((\mE-I_D)^{l-1})}} \;,
\end{equation}
$I_D$ standing for the identity matrix.
Note the non-standard normalization factor $\Lf{((\mE-I_D)^{l-1})}$, which results from the absence of transitions $i \rightarrow i$ in the structure chain $\pC$. The diagonal elements of $\mE$, describing these transitions, thus have to be withdrawn.

Let us emphasize that the maximal length $\lmax$ is at most $D$, but it can be less than $D$ if there are null coefficients on the upper
part of $\mE$. For instance, if we consider the $4$-dimensional matrix 
\begin{equation}
\mE = 
\begin{pmatrix} 
1 & 1 & 1 & 0 \\
0 & 1 & 0 & 1 \\
0 & 0 & 1 & 1 \\
0 & 0 & 0 & 1 
\end{pmatrix} \; ,\; \mA_{ij} = 1
\end{equation}
one has $\lmax=3$ and the structure chains of maximal length are
\begin{equation}
 \pC = (1,2,4), \quad  \pC = (1,3,4).
\end{equation}

\subsection{Limit distributions}
Injecting the distribution $Q(\nu)$ given in \Eqref{eq:FDlim} into \Eqref{eq:Psi-s} yields the limit distribution for the law of large numbers
\begin{equation} \label{eq:lim:LLN}
\Psi(s) =  \sum_{\pC, \len{\pC }=\lmax} p(\pC|\lmax) \int \lU_\pC (\nu)
\delta  \p{ s - \sum_{i,j=1}^D \nu_{ij} \mu_{ij} } \prod_{i,j=1}^D d\nu_{ij} \; .
\end{equation}
Keeping only nonzero frequencies $\nu_{ij}$ (see the definition \Eqref{def:manifold-MC} of the manifold $\Sp_\pC$) and relabelling them as $\alpha_k$, $k=1,\dots,\lmax$, \Eqref{eq:lim:LLN} can be rewritten in a simpler way as
\begin{multline} \label{eq:lim:LLN2}
\Psi(s) =  \sum_{\pC, \len{\pC }=\lmax } (\lmax-1)! \; p(\pC|\lmax)  \\ 
 \int \delta \p{ \sum_{k=1}^{\lmax} \alpha_k -1 }
\delta \p{ s - \sum_{i,j=1}^{\lmax} \alpha_k  \mu_{\pC_k \pC_k} } \prod_{k=1}^{\lmax} d\alpha_k \; ,
\end{multline}
where the integral is over the domain ${[0,1]^\lmax}$.
Although the integral in \Eqref{eq:lim:LLN2} has a rather complicated expression, it contains only constant factors and Dirac distributions, so that
it is possible to recast it as the volume of a particular manifold.
Using this geometric interpretation, it can be shown that $\Psi (s)$ is a piecewise polynomial in $s$.
Section~\ref{sec:algo} briefly explains this result and presents an exact algorithm to compute explicitly the
limit distribution given in~\Eqref{eq:lim:LLN2}.

Similarly, the limit distribution for the central limit theorem can be derived by combining \Eqref{eq:FDlim} with \Eqref{eq:Phi-z}, leading to

\begin{equation} \label{eq:lim:CLT}
\Phi (z) =  
\sum_{\pC , \len{\pC }=\lmax } p(\pC|\lmax) \int \frac{\lU_\pC(\nu)}{\sqrt {2\pi \sum_{i,j=1}^D \nu_{ij} \sigma_{ij}^2}} 
e^{- z^2 / \left[ 2 \sum_{i,j=1}^D \nu_{ij} \sigma_{ij}^2 \right] } \prod_{i,j=1}^D d\nu_{ij} \; ,
\end{equation}
which simplifies to
\begin{multline} \label{eq:lim:CLT2}
\Phi (z) =  
\sum_{\pC , \len{\pC }=\lmax } (\lmax-1)! \; p(\pC|\lmax) \\  \int
\frac{\delta \p{ \sum_{k=1}^{\lmax} \alpha_k -1}}{\sqrt {2 \pi\sum_{k=1}^{\lmax} \alpha_k \sigma_{\pC_k \pC_k}^2}} \,
e^{- z^2 / \left[ 2 \sum_{k=1}^{\lmax} \alpha_k \sigma_{\pC_k \pC_k}^2 \right] } \prod_{k=1}^{\lmax} d\alpha_k \;.
\end{multline}
Unfortunately, we were not able to obtain a simpler and more explicit expression for this limit distribution in the generic case --see however \cite{Angeletti2013:MatSum:Letter} for a simple example.

Eqs.~\eqref{eq:lim:LLN2} and \eqref{eq:lim:CLT2} establish that non-standard limit distributions emerge in presence of
irreversible transitions. For both the law of large numbers and the central limit theorem, these non-standard limit distributions are
discrete mixtures of continuous mixtures of the associated standard distributions. More precisely, continuous
mixtures appear if $\lmax>1$, and discrete mixtures emerge in presence of multiple paths of maximal length $\lmax$.
Note that the results obtained in \cite{Angeletti2013:MatSum:Letter} (apart from the ergodic case corresponding to irreducible matrices $\mE$, that is, one single block in the Perron-Frobenius decomposition \eqref{eq:Perron}) are recovered in the limiting cases $\lmax=1$ and $\lmax=D$. For $\lmax=1$, all structure chains $\pC$ have length one (the matrix $\mE$ is diagonal), and the limit distributions \eqref{eq:lim:LLN2} and \eqref{eq:lim:CLT2} are discrete mixtures (the integral cancels out due to the delta distribution). For $\lmax=D$, there is only one structure chain of length $D$, so that only the continuous mixture remains in \eqref{eq:lim:LLN2} and \eqref{eq:lim:CLT2}, in agreement with the results of \cite{Angeletti2013:MatSum:Letter}.

In summary, Eqs.~\eqref{eq:lim:LLN2} and~\eqref{eq:lim:CLT2} fully characterize the limit distributions of the sum $S(\vec X)$ in the case of totally irreversible structure matrices, i.e. whenever the inner structure of the block $\mB{c}$ is trivial.

\section{Reduction to totally irreversible models} \label{sec:reduced}

We shall now characterize the limit distribution of $\S(\vec X)$ in the presence of
non-trivial structures for the blocks $\mB{c}$. As stated in \cite{Angeletti2013:MatSum:Letter}, 
if the matrix $\mE$ itself is irreducible, then the sum $\S(\vec X)$ converges to a standard limit distribution. 
A natural conjecture at this point is that the inner structure of the block $\mB{c}$ does not influence the limit distribution
of $\S(\vec X)$. 
The justification of this conjecture, given below, is quite technical but relies on three main ideas which can be summarized as follows
\begin{enumerate}
\item The time spent inside a block $\mB{c}$ with $\Lambda_c < \Lambda$ (see Eq.~(\ref{eq:max:Lambda})) is almost surely negligible; \label{it:dom}
\item Inside a dominant block $\mB{c}$ with $\Lambda_c = \Lambda$, the Markov chain $\Gamma $ is asymptotically homogeneous and converges rapidly to its steady state; \label{it:conv}
\item The dominant class chain $\pdC$, obtained by removing non dominant classes $c$ from $\pfC$, is  equivalent to 
the hidden Markov chain of a reduced model ($\mAr,\mEr)$, where $\mEr$ is a totally irreversible matrix. \label{it:class}
\end{enumerate}

The first point implies that only the dominant classes play an important role in the statistics of $\S(\vec X)$.
The second point states that the fine dynamics of $\Gamma $ inside a block is irrelevant to the statistics of $\S(\vec X)$.
Inside a given block $\mB{c}$,  we can replace $\Gamma $ by an 'averaged' $\iid$ random variable with distribution $\mP_{c}$ without modifying 
the limit distribution of $\S(\vec X)$.
Combining these two points shows us that the hidden Markov chain level described by $\Gamma $ contains too much details for our needs. The information which really matters 
is already available on the coarser class level described by $\pdC$.
The third point then shows that this dominant class chain $\pdC$ can be reinterpreted 
as the hidden Markov chain of a specific random vector $\vec Y$ with matrix representation, associated to a totally irreversible matrix $\mEr$.
By combining these three points together, we can construct a totally irreversible model $\vec Y$ such that
\begin{equation}
\begin{cases}
 \S( \vec X) \Dist{\sim} \S(\vec Y) \text{ when } N \rightarrow \infty,\\
 \vec Y \text{ is a totally irreversible model.}
\end{cases}
\end{equation}
At an intuitive level, $\vec Y$ is the random process obtained by 'forgetting'
the inner structure of the blocks $\mB{c}$. We are constructing a process evolving
at the class level $\pdC$ rather than the state level $\Gamma $. Once this reduced model constructed,
we can obtain the limit distribution of $\S(\vec X)$ by applying to $\S(\vec Y)$ the results obtained in section \ref{sec:tir} for totally irreversible models. 
The different steps of this reasoning are presented in details below.

\subsection{Dominant classes}
From Eq.~\eqref{eq:cstructure}, we know that the chain $\Gamma $ jumps from irreducible classes to irreducible classes.
The first step in our reasoning is to evaluate the relative 'time' $t_c$ spent by $\Gamma $ inside a class $\Xi_c$ before jumping to the next irreducible class:
\begin{equation} \label{eq:def-tc}
t_c(\Gamma) = \frac{1}{N+1}\; \card \Set{ k | \pfC_k = c } \; ,
\end{equation}
where $\pfC$ is the chain of class defined in \Eqref{eq:cshape}.
To evaluate the distribution of $t_c(\Gamma)$, let us first introduce $n=\card \Set{ k | \pfC_k = c }$, and consider a chain $\Gamma$ satisfying the constraints
\begin{equation} \label{eq:Gamma-constraints}
\begin{aligned}
\pfC_k &<c \quad \text{for} \quad k=1,\dots,s,\\
\pfC_k &=c \quad \text{for} \quad k=s+1,\dots,s+n,\\
\pfC_k &>c \quad \text{for} \quad k=s+n+1,\dots,N+1.
\end{aligned}
\end{equation}
If a chain $\Gamma$ satisfies the constraints \eqref{eq:Gamma-constraints},
its probability can be written as, using \Eqref{eq:kappa},
\begin{equation} \label{eq:pGamma-constr}
\Prob(\Gamma) = \fLf{\mA_{\Gamma_1 \Gamma_{N+1}}
\left(\prod_{k=1}^s \mE_{\Gamma_k \Gamma_{k+1}} \right)
\left(\prod_{k=s+1}^{s+n-1} \imB{c}_{\Gamma_k \Gamma_{k+1}} \right)
\left(\prod_{k=s+n}^N \mE_{\Gamma_k \Gamma_{k+1}} \right)}
\end{equation}
where $\imB{c}$ denotes the $D\times D$ matrix obtained by setting to zero all coefficients of $\mE$ outside of the block $\mB{c}$.
Note that products are ordered in \Eqref{eq:pGamma-constr}, meaning that for instance $\prod_{k=1}^s \mE_{\Gamma_k \Gamma_{k+1}}$ has to be interpreted as
$\mE_{\Gamma_1 \Gamma_2}\mE_{\Gamma_2 \Gamma_3}\dots \mE_{\Gamma_{s} \Gamma_{s+1}}$.

Then the probability $P(s,n|c)$ that a chain $\Gamma$ satisfies the constraints \eqref{eq:Gamma-constraints} is obtained by summing $\Prob(\Gamma)$, as given in \Eqref{eq:pGamma-constr}, over all chains $\Gamma$ compatible with the constraints. In order to do so, it is easier to split the chain $\Gamma$ into $5$ parts $(\Gamma_1,\dots,\Gamma_{s})$, $\Gamma_{s+1}$, 
$(\Gamma_{s+2}\dots \Gamma_{s+n-1})$, $\Gamma_{s+n}$ and $(\Gamma_{s+n+1}, \dots, \Gamma_{N+1})$. Then, we can subdivide accordingly the matrix $\mE$ into
\begin{equation}
 \mE = \begin{pmatrix} 
\mE_{<c} & \mE_{\rightarrow c} & *                 \\
0       & \mB{c}             & \mE_{c\rightarrow} \\
0       & 0                  & \mE_{>c}           \\  
   \end{pmatrix}.
\end{equation}
The submatrices $\mE_{<c}$ and $\mE_{>c}$ are square matrices of respective dimensions $\sum_{c'<c} D_{c'}$ and $\sum_{c'>c} D_{c'}$. 
With this notation, summing over all chains $\Gamma$ satisfying the constraint \eqref{eq:Gamma-constraints} yields 
\begin{equation}
P(s,n|c) = \fLf { \mE_{<c}^{s-1} \, \mE_{\rightarrow c} \, \mB{c}^{n-1} \, \mE_{c \rightarrow } \, \mE_{>c}^{N-s-n} } \;.
\end{equation}
The probability distribution of the relative time $t_c$ introduced in \Eqref{eq:def-tc} is then obtained by summing over $s$, and using the relation $n=Nt_c$,
\begin{equation} \label{eq:tdist}
p (t_c = t ) =  \sum_{s=0}^{N(1-t)} \fLf { \mE_{<c}^{s-1} \, \mE_{\rightarrow c} \, \mB{c}^{n-1} \, \mE_{c \rightarrow } \, \mE_{>c}^{N-s-n} } \;.
\end{equation}
Since $\mE_{<c}$ and $\mE_{>c}$ are block diagonal submatrices, the spectrum of these matrices is a subset of the spectrum of $\mE$.
The matrix powers in \Eqref{eq:tdist} can therefore be approximated, for large values of the exponents, by
\begin{equation}
  \label{eq:tc:dom_approx}
\begin{aligned}
||\mE_{<c}^s||  &\le  e^{s \ln \Lambda + O(\ln S)}, \\
||\mE_{<c}^{N-s-n}||  &\le  e^{(N-s-n) \ln \Lambda + O(\ln (N-s-n) )  }, \\
||\mB{c}^{n-1}||  &=  e^{(n-1) \ln \Lambda_c + O(\ln n) }.
\end{aligned}
\end{equation}
To leading exponential order in $N$, the probability distribution of $t_c$ is then dominated by a term depending only on the dominant eigenvalue $\Lambda_c$ of $\mB{c}$
\begin{equation} \label{eq:tdist:1bis}
p (t_c = t ) \le e^{ - N t [ \ln \Lambda - \ln \Lambda_c ] + O(\ln N) } \;.
\end{equation}
Consequently, if $\Lambda_c< \Lambda$, the probability of spending a relative time $t_c>0$ inside the block $\mB{c}$ decreases exponentially
with $N$. This exponential decay implies that the probability of observing a value $t_c$ greater than $t$ is less than
\begin{equation} \label{eq:tdist2}
p (t_c > t ) \le e^{ - N t [ \ln \Lambda - \ln \Lambda_c ] + O(\ln N) }\;.
\end{equation}
For a given $N$, the probability of observing a value $t_c$ greater than $\tcUpBl$ is bounded according to
\begin{equation}
\label{eq:t_c:sup_bound}
p\left(t_c > \tcUpB \right) \le e^{- \NcUpB [\ln \Lambda - \ln \Lambda_c] + O(\ln N)}.
\end{equation}
If we observe realizations of $S(\vec X)$ for increasing values of N, then
\begin{equation}
\label{eq:t_c:inf_sum}
\sum_{N=1}^{+\infty} p \left(t_c > \tcUpB \right) < \infty.
\end{equation}
In this situation, the Borell-Cantelli lemma states that almost surely the event $\left( t_c > \tcUpBl \right)$ happens
only a finite number of times. In other words, almost surely for $\Lambda_c < \Lambda$ and $N$ large enough 
\begin{equation} \label{eq:tc_nond}
t_c \as{\le} \tcUpB, \qquad N \rightarrow +\infty \; .
\end{equation}
The total time $N t_c$ spent inside a non-dominant block $\mB{c}$ increases very slowly with $N$. 
The contribution of these non-dominant blocks to the global transition frequencies $\nu_{i,j}$ is negligible.
More precisely, if we call $\bnu{c}$ the transition frequencies inside the block $\mB{c}$, defined as
\begin{equation}
\bnu{c}_{ij} = \frac{1}{(N+1) t_c} \; \card \Set{ l |  \Gamma_l=i, \Gamma_{l+1}=j , (i,j) \in  \sB{c} }  
\end{equation}
(here, $t_c>0$ by definition, because the block $\mB{c}$ is assumed to be visited)
then in the limit $ N\rightarrow  +\infty $
\begin{equation} \label{eq:nu:block}
\vec{\nu }_{ij}  \approx   \sum_{c} t_c \bnu{c}_{ij} \; .
\end{equation}
In \Eqref{eq:nu:block}, the transitions between blocks have been eliminated since there are at most $D$ transitions
between blocks, which are thus negligible in the limit $ N \rightarrow  \infty $. In a similar way,
for a non-dominant block with $\Lambda_c<\Lambda$, \Eqref{eq:tc_nond} implies that
$t_c \bnu{c}_{ij} \as{\rightarrow} 0$. Consequently, \Eqref{eq:nu:block} can be further
simplified to
\begin{equation} \label{eq:nu:block:d}
\vec{\nu }_{ij} \as{\approx }  \sum_{c,\, \Lambda_c = \Lambda } t_c  \bnu{c}_{ij} \;.
\end{equation}
Only the dominant blocks $\mB{c}$ contribute meaningfully to the statistics of the sums. 
If we consider only the chains for which the relative time $t_{c}$ spent inside a dominant block $\mB{c}$
grows proportionally to $N$ when $N \rightarrow  \infty $, then the limit distribution of $\nu $ can be directly derived from the limit distributions of
$\nu({c})$ and $t_{c}$. This is interesting because $\Gamma $ is a Markov chain and satisfies the strong Markov property. More precisely, 
if we call $\sPth{c}$ the subchain of $\Gamma $ such that for all $l$, $\sPth{c}_l \in  \Xi_c$ then $\sPth{c}$ is the hidden Markov chain
associated with the structure matrix $\mB{c}$ and the probability density matrix $\mbP{c}$ obtained by restraining $\mP$ to
the indices $\Xi_c$. The form of the matrix $\mA$ is more complex and depends on other parts of the chain $\Gamma$, and not only on the subchain $\sPth{c}$.
However, this matrix $\mA$ does not play any role in the determination of the limit distribution of $\nu({c})$ as we will see 
in the next section.

\subsection{Convergence inside dominant irreducible blocks}
As mentioned before, one of the main difficulties hindering the derivation of
a limit distribution for $\nu $ is the inhomogeneous nature of the Markov chain $\Gamma $.
If the hidden Markov chain was homogeneous, it would be relatively easy to characterize
the behavior of $\Gamma $ and consequently $\nu$. In our context, in most cases, the 
hidden Markov chain is inhomogeneous. However, it can be shown that inside an dominant irreducible block $\mB{c}$
the Markov chain $\sPth{c}$ is asymptotically homogeneous.

If we suppose that $\mB{c}$ is aperiodic, then there is only one eigenvalue with maximal modulus $\Lambda_c$.
We consider a dominant block $\mB{c}$ for which $\Lambda_c=\Lambda$ 
The power of  $\mB{c}$ can be approximated at large $n$ by
\begin{equation} \label{eq:Bcn}
\mB{c}^n \approx  \Lambda^n \vEig{c} \fEig{c}^T
\end{equation}
where $\vEig{c}$ and $\fEig{c}$ are the positive right- and left-eigenvectors associated to $\Lambda_{c}$, normalized as $\sum_i \vEig{c}_i=1$ and 
\begin{equation} \label{eq:eigen_norm}
\fEig{c}^T \vEig{c}=1.
\end{equation}
We now show that far enough from the end point, the Markov chain $\sPth{c}$ is approximately homogeneous. Denoting its length as $n$,
we first compute the transition rate for $k < n - \sqrt{n}$ and $n \rightarrow  + \infty $, as
\begin{equation} \label{eq:asympt:hom}
\begin{aligned}
p(\sPth{c}_{k+1}= j| \sPth{c}_{k}=i, \, \sPth{c}_{n+1} = f ) &= \mB{c}_{ij} \frac{\mB{c}^{n-k}_{jf}}{\mB{c}^{n-k+1}_{if}}  \\
&\approx  \mB{c}_{ij} \frac{\vEig{c}_j }{\Lambda\vEig{c}_i} \; ,\\
\end{aligned}
\end{equation}
where we have used \Eqref{eq:Bcn} in the second line of \Eqref{eq:asympt:hom}.
The transition rate in Eq.~(\ref{eq:asympt:hom}) is now independent of $k$ and of $f$, so that we shall simply denote it as $p(i \rightarrow j|c)$, emphasizing the block dependence.
It can be verified that the transition probabilities are normalized
\begin{equation}
\sum_{j=1}^{D_c} p (i \rightarrow j|c) = \sum_{j=1}^{D_c}  \mB{c}_{ij} \frac{\vEig{c}_j }{\Lambda  \vEig{c}_i}  = \frac{\Lambda \vEig{c}_i} {\Lambda  \vEig{c}_i} =1 \; .
\end{equation}
It is therefore possible to use standard results for homogeneous Markov chains \cite{Seneta06} to prove that the chain converges to its stationary state $\vStat{c}$. This stationary state can be expressed
using the left and right dominant eigenvectors of $\mB{c}$, $\fEig{c}$ and $\vEig{c}$, as
\begin{equation}
\vStat{c}_i = \vEig{c}_i \, \fEig{c}_i .
\end{equation} 
Indeed, we have
\begin{equation} 
 \sum_{i=1}^{D_c} \vStat{c}_i \, p(i \rightarrow j|c)  
=   \sum_{i=1}^{D_c} \vEig{c}_j \, \frac{\fEig{c}_i \mB{c}_{ij}} {\Lambda }
=  \vStat{c}_j .
\end{equation}
Note that the normalization of $\vStat{c}$ derives from the normalization of the eigenvalues $\vEig{c}$ and $\fEig{c}$ chosen in \Eqref{eq:eigen_norm}. \Eqref{eq:eigen_norm} can therefore be interpreted as imposing  that $\vStat{c}$ is a discrete probability distribution.
Moreover, convergence theorems for homogeneous Markov chains \cite{Seneta06}
state that the convergence
speed is exponential with a finite time scale $\tau$.  Beyond this time scale, for instance for $k \in  [\tau \sqrt{N}, N t_c -\tau \sqrt{N}]$,
$\sPth{c}$ can be considered to be in its stationary state.
Consequently, the relative time spent at the stationary distribution tends to $t_c$ when $N t_c \rightarrow \infty $.
Inside the block $\mB{c}$, the transition frequencies $\bnu{c}$ converge to a non-random limit,
\begin{equation} \label{eq:nup:hom}
\bnu{c}_{ij} \convas{N} \bar\nu(c)_{ij} \equiv \frac{1}{\Lambda}\, \fEig{c}_i \, \mB{c}_{ij} \, \vEig{c}_j \; .
\end{equation}
A key consequence of \Eqref{eq:nup:hom} is that the matrix $\mA$ does not play any
role in the limit distribution of $\bnu{c}$. 
It is important to note that \Eqref{eq:nup:hom} is valid even for an irreducible periodic matrix $\mB{c}$ with period $P$.
A proof is presented in Appendix~\ref{app:periodic}.
Briefly, it relies on the fact that each of the $P$ subchains
\begin{equation}
\PGamma c o = (\sPth{c}_{o}, \dots, \sPth{c}_{o+ kP}, \dots, \sPth{c}_{o + P \floor{ (N t_c-o) / P }}), \qquad o=1,\dots,P,
\end{equation}
does converge to a steady state ($\floor{x}$ denotes the integer part of $x$). This steady state depends on $\mA$.
Nevertheless, the global transition frequencies $\vec{\nu }$ are the average of the transition frequencies
of $\PGamma c 1,\dots, \PGamma c P$ and do not depend on $\mA$. Moreover, these averaged transition frequencies also satisfy
\Eqref{eq:nup:hom}. In terms of sum statistics, the periodic case is therefore equivalent to the simpler aperiodic case.

The convergence, inside a given block $\mB{c}$, of the transition frequencies
$\bnu{c}_{ij}$ to non-random values $\bar\nu(c)_{ij}$ implies that the sum of non-identically distributed variables $\cX{ij}_k$, with 
$i,j \in \sB{c}$ can be replaced by a sum of identically distributed 'averaged' variables $Y_{c,k}$ of distribution $\mP_c(y)$, defined 
by its characteristic function $\varphi_c(q) \equiv \int_{-\infty}^{\infty} \mP_c(y) \, e^{iqy} dy$, given by
\begin{equation} \label{eq:def:chiY}
\varphi_c(q) = \prod_{i,j \in \sB{c}} \varphi_{ij}(q)^{\bar\nu_{ij}(c)} \;,
\end{equation}
where $\varphi_{ij}(q) \equiv \int_{-\infty}^{\infty} \mP_{ij}(x) \, e^{iqx} dx$ is the characteristic function of the variable $\cX{ij}$.
If we consider a given vector $t$ of relative times with $ t_c =0$ if $\Lambda_c < \Lambda$, then  
the sum $S(\vec X| t )$ conditioned on $t$ becomes equivalent in distribution to a 
sum over dominant blocks of variables $Y_{c,k}$
\begin{equation} \label{eq:sum-XY}
S(\vec X| t ) \Dist{ \approx } \sum_{c, \Lambda_{c}=\Lambda}
\sum_{k=1}^{Nt_{c}} Y_{c,k}.
\end{equation}
The relative time $t_c$ is now playing a role similar to that of the frequencies $\nu_{ij}$ in \Eqref{eq:sum:rew}.
For some characteristic functions $\varphi_{ij}$, the function $\varphi_c$ defined
by \Eqref{eq:def:chiY} might not be semidefinite positive. In this situation, the inverse Fourier transform of $\varphi_c$
is non positive and $\varphi_c$ is not the characteristic function of any valid probability distribution.
Nevertheless, we can avoid this difficulty if we approximate $\varphi_c$ by the characteristic function
of a gaussian random variable of same mean and variance
\begin{equation}
 \varphi_c(q) \approx \exp \p{  \varphi_c'(0) q + (\varphi_c''(0) - \varphi_c'(0)^2) \frac{q^2}{2} }.  
\end{equation}
Due to the central limit theorem, \Eqref{eq:sum-XY} is still valid when $\varphi_c$ is replaced
by this approximate gaussian variable.

Building on \Eqref{eq:sum-XY}, we shall see below how to make a more precise mapping to a totally irreversible model based on the variables $Y_{c,k}$, where the chain of dominant classes plays the role of the Markov chain $\Gamma$.

\subsection{Reduced model}

We have seen above that it is possible to forget the internal
structure of $\mB{c}$.
We now proceed to construct a class-level model which retains all the information necessary to describe the limit distribution of $S(\vec X)$.

A first important point is that we already know that for a non-dominant block $\mB{c}$ with $\Lambda_c < \Lambda $, almost surely
$t_c \rightarrow  0$. 
Starting from the chain of classes $\pfC$, we can construct the chain of dominant blocks $\pdC$
by removing the non-dominant classes, e.g.
\begin{equation}
\pdC= \cancel{ \pfC_{1} }\; \pfC_2\; \cancel{\pfC_3} \; \cancel{\pfC_4} \; \pfC_5 \; \pfC_6 \; \dots \pfC_{N+1} .
\end{equation}
The length $\len \pdC$ of the chain $\pdC$ is no longer fixed to $N$. More precisely, if we define $m_c$ as the time
spent transitioning between the $(c-1)$-th and $c$-th distinct dominant classes, we have 
\begin{equation}
N=\len \pdC + \sum_{c=1}^{\len m} m_c.  
\end{equation}
Note that by convention, we call $m_1$ the time spend before arriving to the first dominant class and
$m_{\len m}$ the time remaining after the last dominant class.
Since $t_c \as{<} \tcUpBl $ for non-dominant class $c$ with $\Lambda_c<\Lambda $, we have an almost sure upper bound on $m_c$
\begin{equation} \label{eq:m_c:ub}
m_c \as{<} D \tcUpB.
\end{equation}
This upper bound on $m_c$ translates into a lower bound on $\len \pdC$
 \begin{equation} \label{eq:sh_t}
 N - D^2 \tcUpB \as{<} \len \pdC \leq  N \; .
\end{equation}
We can therefore consider that $\len \pdC \approx  N$.
Using a notation similar to that defined in Sect.~\ref{sec:tir}, we 
consider $\pCr$ the chain of distinct classes appearing in $\pdC$ 
\begin{equation} \label{eq:soc}
\pdC  = (\repeated{\pCr_1}{t_{\pCr_1}}, \dots, \repeated{\pCr_{\len{\pCr }}}{t_{\pCr_{\len{\pCr }}}}).    
\end{equation}
With this definition \eqref{eq:soc}, the chain $\pdC$ is completely equivalent to the couple $(\pCr,t_c)$.
In particular, we can reread $S(\vec X | t)$ as 
\begin{equation} \label{eq:cond:equiv:pdC}
 S( \vec X | t )  \equiv  S(\vec X | \pdC ).
\end{equation} 
In the following, we wish to determine the distribution $p(\pdC)$ of the chain $\pdC$ of dominant classes. To this aim, it is useful to introduce a 'shadow' transition matrix $\mS$ defined as
\begin{equation} \label{eq:shw}
\mS =  \mE - \sum_{c, \, \Lambda_c=\Lambda} \imB{c},
\end{equation}
that characterizes the role of non-dominant blocks in the dynamic of $\pdC$. Such blocks play the role of transient intermediaries between irreducible blocks.
Summing, in \Eqref{eq:kappa}, over all possible chains $\Gamma $ sharing the same dominant class chain $\pdC$ and transition times
$m_c$ leads to
\begin{equation} \label{eq:red:pred}
p(\pdC, m_c ) \as{ = }\fLf {  \p{ \mS^{m_1}  \imB{\pCr_1}^{N t_1} \dotsm \mS^{m_{\len \pCr}}  \imB{\pCr_{\len{\pCr}}}^{N t_{\len{\pdC}}}} \mS^{m_{\len{\pCr}+1}}} \;.
\end{equation}
In \Eqref{eq:red:pred}, the term $\imB{c}^{N t_c}$ corresponds to the contribution of all subchains $\sPth{c}$ of length $N t_c$
whereas the term $\mS^{m_c}$ originates from the subchains transitioning from one dominant block to the next one. 
We can then sum over all $m_c$ to obtain the distribution of $\pdC$. Moreover, \Eqref{eq:m_c:ub} states 
that almost surely the transition times $m_c$ are less than $D \tcUpBl$. It is therefore sufficient 
to sum over all $m_c$ fulfilling this upper bound    
\begin{equation} \label{eq:red:pred:s}
p(\pdC) \as{ = } \sum_{ m_c < D \tcUpBl } 
\fLf { \p{ \mS^{m_1}  \imB{\pCr_1}^{N t_1} \dotsm \mS^{m_{\len \pCr}}  \imB{\pCr_{\len{\pCr}}}^{N t_{\len{\pCr}}} } \mS^{m_{\len{\pCr}+1}}} \;.
\end{equation}
Using in \Eqref{eq:red:pred:s} the expression of the powers of $\mB{c}$ given in \Eqref{eq:Bcn}, one finds
\begin{multline} \label{eq:red:0}
 p(\pdC) \approx  \sum_{m_c \le D \tcUpBl } 
 \fLf{\fracpow{\mS}{\Lambda }{m_1} \vEig{\pCr_1}  \Lambda^{N} \fEig{\pCr_{\len \pCr}}^T \fracpow{\mS}{\Lambda }{m_{\len{\pCr}+1}}} 
  \\ \prod_{c=1}^{\len{\pCr}-1} \fEig{\pCr_c}^T \fracpow{\mS}{\Lambda }{m_{c+1}}  \vEig{\pCr_{c+1}} 
\end{multline} 
Since the transition matrix $\theta$ contains only non-dominant blocks, its operator norm is strictly inferior to $\Lambda$.
The sum 
\begin{equation} 
\mSl = \sum_k \fracpow{\mS}{\Lambda }{k}
\end{equation}  is therefore convergent. Injecting this limit into \Eqref{eq:red:0} yields
\begin{equation} \label{eq:red:1}
p(\pdC) \approx  \frac{\Lambda^{N}}{\Lf\p{\mE^N}} \Lf\p{\mSl \vEig{\pCr_1} \fEig{\pCr_{\len \pCr}}^T \mSl } \prod_{c=1}^{\len{\pCr}-1} \left[ \fEig{\pCr_c}^T  \mSl \vEig{\pCr_{c+1}} \right] \;.
\end{equation}
Except for the normalization factor $\Lambda^N/\Lf\p{\mE^N}$, all factors appearing in \Eqref{eq:red:1}
depend only on pairs of dominant classes. This structure is quite remarkable and can be used to construct 
'reduced' matrices of dimension equal to the number of dominant classes $\redc{D}$. In particular, we can define 
reduced structure matrix $\mEr$ and projection matrix $\mAr$, associated to a linear form $\Lfr$
\begin{equation} \label{eq:Er}
\mEr_{c c'} = \fEig{c}^T  \mSl \vEig{c'},
\end{equation}  
\begin{equation} \label{eq:Ar}
\mAr_{c c'} = \Lf\p{ \mSl \vEig{c} \fEig{c'}^T \mSl} , \quad \Lfr\p{M}= \tr\p{ \mAr[T] M} \;.
\end{equation} 
With these definitions, \Eqref{eq:red:1} becomes
\begin{equation} \label{eq:Gen:Prob:Sq:2}
p(\pdC)  \approx  \frac{\mAr_{\pCr_1 \pCr_{\len{\pdC}}
}}{\Lfr\p{\mEr[\len{\pdC}-1]}}   \prod_{c=1}^{\len{\pCr}-1} \mEr_{\pCr_c \pCr_{c+1} } .
\end{equation}
Comparing \Eqref{eq:Gen:Prob:Sq:2} with \Eqref{eq:kappa},
one sees that the probability distribution of the chain $\pdC$ maps to the distribution of a hidden Markov Chain $\pthr$ associated to the matrices $\mEr$ and projection matrix $\mAr$ defined in Eqs.~\eqref{eq:Er} and \eqref{eq:Ar}. A minor remark is that the matrix
$\mAr$ is necessarily non zero due to the assumption that all classes $\sC{c}$, and in particular all dominant classes, were reachable
for the original pair of matrices $(\mE,\mA)$. 

In addition, the shape of $\mEr$ is severely constrained. First, the eigenvectors $\vEig c$ and $\fEig c$ have non-zero coefficients only for indices $ i \in \sC{c}$.
Similarly, the matrix $\mSl$ inherits from $\mE$ its upper block triangular structure:
\begin{equation} \label{eq:Sl:shape}
 \mSl_{i,j} > 0 \implies \exists c\le c', \quad (i,j) \in \sC{c}\times\sC{c'} .
\end{equation}
Combined with the shape of the eigenvectors $\vEig{c}$ and $\fEig{c'}$, \Eqref{eq:Sl:shape} yields 
\begin{equation}
\mEr_{cc'}>0 \implies c \le c'.
\end{equation}
The matrix $\mEr$ is therefore a upper triangular matrix.
Moreover, inside dominant block $\mB{c}$, the transition matrix $\mSl$ is equal to the identity matrix
\begin{equation}
 \p{ \mSl_{i ,j } }_{i,j \in \sC{c}} = \mId .
\end{equation}
Taking in account the normalization of the eigenvectors \eqref{eq:eigen_norm}, we have on the 
diagonal of $\mEr$
\begin{equation}
\mEr_{cc} = \fEig{c}^T  \vEig{c'} = 1.
\end{equation}
Since all the diagonal coefficients are equal, the upper triangular matrix $\mEr$ is a totally irreversible matrix.
Consequently, \Eqref{eq:Gen:Prob:Sq:2} is formally equivalent to a version of \Eqref{eq:pGamma-irrev} where 
$N$ has been replaced by $(\len{\Gamma}-1)$.

Combining the distribution $p(\pdC)$ given in \Eqref{eq:Gen:Prob:Sq:2} with the reformulation in terms of effective variables $Y_{c,k}$ proposed in \Eqref{eq:sum-XY}, we arrive at a full characterization of a reduced model.
To this aim, we first define the matrix of probabilities $\mPr_{c c'}(y)$ as
\begin{equation}
  \label{eq:Pr}
  \mPr_{c c'} (y) =
\begin{cases}
\mP_c(y) & \text{if } c=c' \;, \\
\delta(y) & \text{otherwise} \;.
\end{cases} 
\end{equation}
The diagonal distributions $\mPr_{c c}(y)$ correspond to the averaged distribution inside the dominant block $\mB{c}$ introduced in \Eqref{eq:def:chiY}.
Non-diagonal distributions are arbitrary, and we have chosen them as Dirac distributions so that they bring strictly no contribution to the sum $S$.
Another choice for the non-diagonal distributions $\mPr_{c c'}(y)$ would only modify finite-$N$ effects, and not the limit distribution of the sum $S$.
We further define the matrix function $\mRr(y)$ as
\begin{equation}
\mRr_{c c'}(y) = \mEr_{c c'} \mPr_{c c}(y) \;.
\end{equation}
Then the random vector $\vec Y$ is defined by the following joint probability distribution, analogous to \Eqref{eq:Def},
\begin{equation}
P^*(y_1,\dots,y_N) = \frac{1}{\Lfr (\mEr[N])}\, \Lfr \p{ \mRr(y_1) \mRr(y_2) \dots \mRr(y_N) } \;.
\end{equation}
By construction, we have for the hidden chain $\pthr$
\begin{equation} \label{eq:red:hidden_eq}
p_X( \pdC ) \approx p_Y( \pthr) .
\end{equation}
From the definition \Eqref{eq:Pr} of the probability distributions $\mPr_{c,c'}(y)$, the sum $S(\vec Y|\pthr)$ reduces exactly to a sum over diagonal terms 
\begin{equation} 
S(\vec Y|\pthr) = \sum_{c, \Lambda_{c}=\Lambda}
\sum_{k=1}^{Nt_{c}} Y_{c,k} \;.
\end{equation}
Therefore, Eqs.~\eqref{eq:sum-XY} and~\eqref{eq:cond:equiv:pdC} imply that 
\begin{equation} \label{eq:red:cond_eq}
S(\vec X|\pdC) \Dist{ \approx } S(\vec Y|\pthr) \;.
\end{equation}
We can then combine our results on the hidden Markov chain level from \Eqref{eq:red:hidden_eq} and 
on the conditioned level from \Eqref{eq:red:cond_eq} to obtain
\begin{equation}
 S(\vec X) \Dist{\approx} S(\vec Y).
\end{equation}
We have thus shown that the sum $S(\vec X)$ of a generic matrix-correlated random vector can be mapped onto the sum $S(\vec Y)$ of a random vector belonging to the class of totally irreversible models. The limit distributions for the sum of generic  matrix-correlated random vectors are thus the same as that found for the class of totally irreversible models.
Starting from a given matrix-correlated random vector $\vec X$, the limit distribution is obtained by determining explicitly the associated reduced model.

\section{Algorithmic computation of the limit distributions} \label{sec:algo}
Even though we have characterized the form of the limit distribution for $\S(\vec X)$, the construction of the reduced totally irreversible model $\vec Y$ is quite complex, and the computation of the limit distribution for a generic
matrix representation $(\mA,\mE,\mP)$ is still a non-trivial task.
We propose in this section a brief algorithmic summary of this construction, illustrated by a randomly generated concrete example.
Due to the algorithmic nature of this section, we have made publicly available
\footnote{http://perso.quaesituri.org/florian.angeletti/Softwares/Scientific} 
a set of python scripts which mirror the steps of this construction.

As an example, we consider the following structure matrix $\mEx$ and projection matrix $\mAx$ 
\begin{equation}
\mEx = \begin{pmatrix}
\frac{1}{3}& 0& 0& \frac{1}{2}& \frac{2}{3}& 0\\[0.1cm]

0& \frac{3}{4}& \frac{1}{4}& 0& 0& 0\\[0.1cm]
0& \frac{1}{2}& \frac{1}{2}& 0& 0& 0\\[0.1cm]
0& 0& 0& \frac{1}{2}& 0& \frac{1}{3}\\[0.1cm]
\frac{3}{4}& 0& 0& 0& \frac{1}{4}& 0\\[0.1cm]
0& \frac{1}{3}& \frac{1}{4}& 0& 0& 1
\end{pmatrix}, \quad \mAx_{ij} = 1   .
\end{equation}

\subsection{Strongly connected classes $\Xi_i$}
The first step is to identify the irreducible classes (called strongly connected components in graph theory) of $\mE$.
In order to do so, an interesting method is to compute a connectivity matrix $\Conn$
\begin{equation} 
\Conn \equiv  \sum_{k=0}^{+\infty } (\epsilon \mE)^k = (\mId- \epsilon \mE )^{-1} .
\end{equation}
For a small enough $\epsilon$, the matrix $(\mId- \epsilon \mE )$ is diagonal dominant
and therefore easily inversible.
Then, there is a path from $i$ to $j$ if and only if $ \Conn_{ij} >0$.
This exact inequality could seem to be troublesome for numerical algorithms.
However, for an $\epsilon$ sufficiently small, the Gauss-Jordan elimination algorithm uses only addition of positive term
to construct $\Conn_{ij}$. Moreover, numerical addition\footnote{without overflow} satisfies the property that the sum of two strictly positive number is still strictly positive. We are therefore in one of the rare cases where the exact inequality $\Conn_{ij} > 0$ is meaningful even when using floating point arithmetic. 

Applying this algorithm to $\mEx$ and replacing strictly positive coefficients by a symbol $"+"$ yields 
\begin{equation}
\ex\Conn = \begin{pmatrix}
+& +& +& +& +& +\\
0& +& +& 0& 0& 0\\
0& +& +& 0& 0& 0\\
0& +& +& +& 0& +\\
+& +& +& +& +& +\\
0& +& +& 0& 0& +
\end{pmatrix} 
\;.
\end{equation}
Once the matrix $\Conn$ has been computed, the next step is to determine a relabelling $i \mapsto i'$
leading to the Perron-Frobenius decomposition.
This relabelling can be found in two steps. First, we identify the strongly connected classes of the graph.
If we call $R(i)_k = \Conn_{ik}$ the $i$th row of the connectivity matrix then two indices $(i,j)$ belong to
the same class if and only if $R(i)=R(j)$:
\begin{equation}
 \{ \sC c \} =  R^{-1}\p{ R \{ 1, \dots, D \} } \;.
\end{equation}
Applying this algorithm to $\Conn$ yields
\begin{equation}
\{ \ex{\sC{c}} \} =  \Set{\Set{1,5},\Set{4},\Set{6},\Set{2,3}} \;.
\end{equation}
Finally,  we need to find a ordering $c \mapsto \tilde{c}$
of these components such that
\begin{equation} \label{eq:clOrder}
\forall (i,j) \in \sC{c} \times \sC{e}, \quad \Conn_{ij} > 0 \implies \tilde c \le \tilde e \;.
\end{equation}
A simple way to find this ordering is to start from the set of classes $V_0=\{ \sC{c} \}$.
We can then look at the subset $V_1$ of classes of $V_0$ which have an antecedent among $V_0$.
The difference set $V_0/V_1$ contains the classes which do not have any antecedent class.
As a consequence, if $\sC{c} \in V_0/V_1$ and $\sC{e} \in V_0$ then we know that $\sC{e} \rightarrow \sC{c}$ is impossible.
Here, we note $\Xi_{c} \rightarrow \Xi_{c'}$ if there is a path going from $\Xi_c$ to $\Xi_c'$,  i.e
if the submatrix $\Conn_{i \in \Xi_{c}, j \in \Xi_{c'}}$ is a non-zero matrix.
In other words, we can safely order the classes of $V_0/V_1$ before the classes of $V_1$ and the ordering of the classes
inside $V_0/V_1$ is arbitrary.  
We can then repeat this procedure by defining $V_{n+1}$ as the subset of classes of $V_n$ with
antecedents among $V_n$:
\begin{equation} \label{eq:app:sseq}
V_{n+1} = \Set{ \Xi  \in V_n | \exists \Xi_c \in V_n/\Xi,\,  \Xi_c \rightarrow  \Xi  },
\end{equation}
Note that the cardinal of the set $V_{n+1}$ is always strictly inferior to the cardinal of the set $V_n$
if $V_n$ is not the empty set. Moreover, there cannot be a chain of distinct classes of length greater 
than $\lmax$.  We have therefore $V_\lmax= \emptyset$ and a finite sequence 
\begin{equation}
V_0 \sqsupset V_1 \sqsupset \dots \sqsupset V_\lmax = \emptyset.  
\end{equation}
We can then partition $V_0$ into the difference sets $U_n$
\begin{equation}
  U_n = V_{n+1}/V_n.
\end{equation}
By construction, if $\sC{c} \in U_l$ and $\sC{c'} \in U_{l'}$ then 
\begin{equation}
  \label{eq:app:vseq:prop}
  \sC{c} \rightarrow \sC{c'} \implies l < l'.
\end{equation}
The sequence $U_n$ defines an ordering of the classes $\Set{\sC{c}}$ which is compatible with 
\Eqref{eq:clOrder}. However, this ordering is only a partial ordering of $\Set{\sC{c}}$.
There may be many total orderings of the indices $i$ compatible with this preorder of the classes, but these different 
orderings are equivalent for our purpose. In our example, we have
\begin{equation} \begin{aligned}
U_0&=\Set{\Set{1,5}} \\
U_1&=\Set{\Set{4}}   \\
U_2&=\Set{\Set{6}}   \\
U_3&=\Set{\Set{2,3}} \\
\end{aligned} \end{equation}
and thus $4$ different potential orderings.
Once a specific relabelling has been found, we obtain the Perron-Frobenius form of the matrix $\mEx$
\begin{equation}
\mEx = \begin{pmatrix}
\frac{1}{3}& \frac{2}{3}& \frac{1}{2}& 0& 0& 0\\[0.1cm]
\frac{3}{4}& \frac{1}{4}& 0& 0& 0& 0\\[0.1cm]
0& 0& \frac{1}{2}& \frac{1}{3}& 0& 0\\[0.1cm]
0& 0& 0& 1& \frac{1}{3}& \frac{1}{4}\\[0.1cm]
0& 0& 0& 0& \frac{3}{4}& \frac{1}{4}\\[0.1cm]
0& 0& 0& 0& \frac{1}{2}& \frac{1}{2}
\end{pmatrix}
\;,
\end{equation}
leading to the following $4$ blocks $\ex{\mB{c}}$:
\begin{equation}
\ex{\mB1} = \begin{pmatrix}
\frac{1}{3}& \frac{2}{3}\\[0.1cm]
\frac{3}{4}& \frac{1}{4}
\end{pmatrix}
\;, \quad
\ex{\mB2} = \begin{pmatrix}
\frac{1}{2}
\end{pmatrix}
\;, \quad
\ex{\mB3} = \begin{pmatrix}
1
\end{pmatrix}
\;, \quad
\ex{\mB4} = \begin{pmatrix}
\frac{3}{4}& \frac{1}{4}\\[0.1cm]
\frac{1}{2}& \frac{1}{2}
\end{pmatrix}
\;.
\end{equation}

\subsection{ Dominant triplet $\p{\Lambda_c, \vEig c, \fEig c}$}
For each block $\mB{c}$, we have to compute the triplet $\Lambda_c, \vEig c, \fEig c$.
Since the blocks are irreducible by definition, the classical power algorithm
can be used directly. This algorithm consists in computing iteratively a vector $v^{(k)}$:
\begin{equation}
v^{(k+1)} = \frac{ \mB{c} v^{k} } {|| \mB{c} v^{(k)}||},
\end{equation}
starting from an initial vector $v^{(0)}=1$.
The vector $v^{(k)}$ converges to the dominant right-eigenvector $\vEig c$ when $k \rightarrow \infty$, and the associated eigenvalue $\Lambda_c$ can be computed as the limit of $\Lambda_c^{(k)}$ for $k \rightarrow \infty$, with
\begin{equation}
\Lambda_c^{(k)} =  \frac{ \p{v^{(k)}}^T \mB{c} v^{(k)} } { \p{v^{(k)}}^T v^{(k)} } \;.
\end{equation}
The same algorithm can be used to compute the dominant left-eigenvector of $\mB{c}$
which is the dominant right-eigenvector of $\mB{c}^T$.

Another possibility is to compute the eigenvalue $\Lambda_c$ by using the characteristic polynomial
of $\mB{c}$. This method is generally a little more amenable to symbolic computations.
For instance, in our example, the diagonal blocks of $\ex\mE$ have been 
constructed to be rational multiples of a stochastic matrix. In this very specific case, it is possible to compute exactly
each triplet and obtain
\begin{equation} \begin{aligned}
 \ex\Lambda_1=1,\quad& \ex{\vEig1}= \begin{pmatrix}
1\\
1
\end{pmatrix},\quad& \ex{\fEig1} = \begin{pmatrix}
\frac{9}{17}\\[0.1cm]
\frac{8}{17}
\end{pmatrix}\\
\ex\Lambda_2=\frac{1}{2},\quad& \ex{\vEig2} = \begin{pmatrix}
1
\end{pmatrix},\quad& \ex{\fEig2} = \begin{pmatrix}
1
\end{pmatrix} \\ 
\ex\Lambda_3 =1,\quad&  \ex{\vEig3} = \begin{pmatrix}
1
\end{pmatrix},\quad&  \ex{\fEig3} = \begin{pmatrix}
1
\end{pmatrix}\\
\ex\Lambda_4=1,\quad& \ex{\vEig4} = \begin{pmatrix}
1\\
1
\end{pmatrix},\quad&  \ex{\fEig4} = \begin{pmatrix}
\frac{2}{3}\\[0.1cm]
\frac{1}{3}
\end{pmatrix}.
\end{aligned}\end{equation} 

\subsection{Limit transition matrix}
We can then identify the dominant and non-dominant blocks. There is however one caveat
here: if the eigenvalues $\Lambda_c$ are computed using a numerical algorithm, exact comparisons between them
could be meaningless. 
However, the convergence condition for the time $t_c$ spent inside a block $\mB{c}$ gives us a natural comparison between
eigenvalues.
\Eqref{eq:tdist2} implies that in order to neglect the time spent inside a block $\mB{c}$, we need to verify that
\begin{equation} \label{eq:neglect}
\fracpow{\Lambda_c}{\Lambda }{N} \ll 1.
\end{equation}
Consequently, \Eqref{eq:neglect} defines a sensible criterion for the comparison between $\Lambda_c$'s.  
With this caveat in mind, we can construct the limit transition matrix $\mS$ from the normalized
structure matrix $\mE/\Lambda $.
\begin{equation}
 \mSl = \sum_{k=0}^{+\infty } \mS^{k} = \p{\mId - \mS }^{-1} \;,
\end{equation}
where $\theta$ is defined in \Eqref{eq:shw}.
In our example,
\begin{equation}
\ex\mSl = \begin{pmatrix}
1& 0& 1& \frac{1}{3}& \frac{1}{9}& \frac{1}{12}\\[0.1cm]
0& 1& 0& 0& 0& 0\\[0.1cm]
0& 0& 2& \frac{2}{3}& \frac{2}{9}& \frac{1}{6}\\[0.1cm]
0& 0& 0& 1& \frac{1}{3}& \frac{1}{4}\\[0.1cm]
0& 0& 0& 0& 1& 0\\[0.1cm]
0& 0& 0& 0& 0& 1
\end{pmatrix}.
\end{equation}

\subsection{Reduced model}
With this, we have obtained all the information needed 
to compute $\p{\mEr,\mAr,\mPr}$ using Eqs.~(\ref{eq:Er}), (\ref{eq:Ar}) and~(\ref{eq:Pr}).
Here, we have for $\ex\mAr$ and $\ex\mEr$
\begin{equation}
\ex\mAr= \begin{pmatrix}
2& 1& 2\\
0& 2& 1\\
0& 0& 2
\end{pmatrix} ,\quad  
\ex\mEr = \begin{pmatrix}
1& \frac{1}{3}& \frac{11}{108}\\[0.1cm]
0& 1& \frac{11}{36}\\[0.1cm]
0& 0& 1
\end{pmatrix} \;.
\end{equation}
We can also determine the path of maximal length, which is unique here, and its probability,

\begin{equation}
\ex{\pC }_1 = [1, 2, 3], \quad p\p{\ex{\pC }_1}=1 \;.
\end{equation}

\subsection{Limit laws for the central limit theorem}

\begin{figure}
\centerline{ \includegraphics[width=0.7\columnwidth]{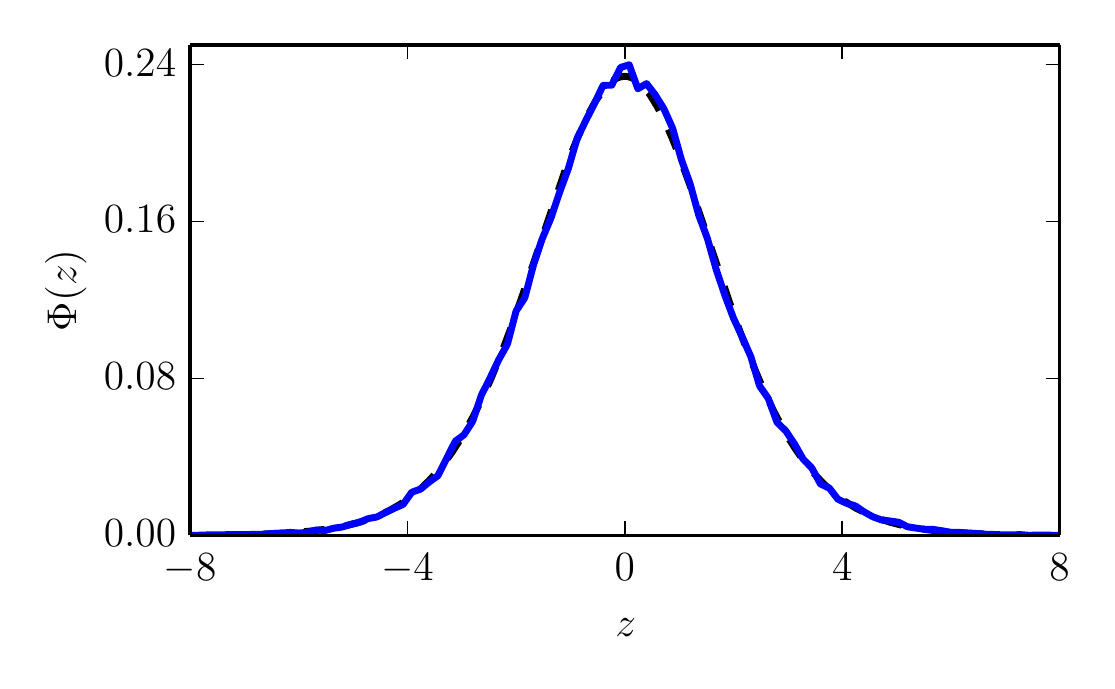} }
\caption{ \label{fig:clt} Limit law $\Phi(z)$ for the centered variable $z=(S-N\mu)/\sqrt{N}$ computed with a Markov integral method. Dashed black line: theoretical distribution; Full blue line:
empirical histogram computed from $100000$ realizations of $\vec X$ with $N=1000$.}
\end{figure}

Once we know the triplet $(\mEr,\mAr,\mPr)$ and the maximal path $\pC $, it is possible to compute the limit distribution $\Phi(z)$
for the central limit theorem using \Eqref{eq:lim:CLT2}. 
If we suppose that the moment matrix $\mQ{1}$ has all its coefficients identical, namely $\mQ{1}_{ij}= \mu$, it is possible to use \Eqref{eq:lim:CLT2}
to compute the limit distribution of the centered variable $z = (S(\vec X) - N\mu )/ \sqrt{N}$.
On the one hand, it does not seem possible to obtain an explicit analytic form for the integral~\eqref{eq:lim:CLT2}.
On the other hand, its form is quite convenient for a Markov integration.
The only difficulty is the presence of the Dirac distribution $\delta(1 - \sum_i \alpha_i )$.
However, in terms of Markov integrals, this distribution corresponds to a uniform sampling of the $\alpha_i$ on the $(\lmax-1)$-simplex
\begin{equation}
\Sp_\pC = \{ \alpha,\quad \forall i ,  \alpha_i >0,\, \sum_{i=1}^{\len{\pC } } \alpha_{\pC_i}=1 \}. 
\end{equation}
Moreover, sampling uniformly on a $\lmax$-simplex can be done by generating $\lmax$
$\iid$ exponential random variables with the same shape parameter and then normalize (accordingly to the $||\cdot ||_1$ norm) the
resulting vector. Fig.~\ref{fig:clt} illustrates the limit law for our example if we choose
the following variance for the reduced model 
\begin{equation}
\sigma^2_{11} = \frac{35}{17}, \quad \sigma^2_{22} =  1.
\end{equation}

\subsection{Limit distribution for the law of large numbers} \label{sec:geom}
In order to determine explicitly the limit distribution $\Psi(s)$ for the sample mean $s=S(\vec X)/N$ for a given structure path $\pC $, we have to evaluate 
the integral~\eqref{eq:lim:LLN2}. There are two essential differences with the case of the central limit theorem. 
First, there is one more Dirac distribution
$\delta \p{ s - \sum_k \alpha_k  \mu_{\pC_k \pC_k} }$. This implies that the integral~\eqref{eq:lim:LLN2} is null except on the manifold
\begin{equation}
K_{\pC }(s) = \Sp_{\pC } \cap H_{\pC }(s) \;,
\end{equation}
with $H_{\pC }(s)$ the hyperplane
\begin{align}
H_{\pC }(s) = \{ \alpha  \in  \R ^{\lmax} | \sum_k \alpha_k \mu_{\pC_k \pC_k} = s \}.
\end{align}
$\Sp_{\pC }$ is the standard $(\lmax-1)$-simplex and enforces the condition that 
the sum of the $\alpha_i$'s is equal to $1$, whereas $H_\pC (s)$ is the set of $\alpha $ corresponding to an average $s$.
Second, except for the Dirac distribution the integral does not contain any varying term.
Consequently, if we restrain the integration domain of \Eqref{eq:lim:LLN2} to the support of the Dirac distribution, we have
\begin{equation} \label{eq:pregeom}
 p \p{ \frac{S(\vec X | \pC )}{n} = s }  = \frac{1}{V}\int_{K_\pC (s)} 1 d\alpha ,
\end{equation}
with $V$ the normalization constant
\begin{equation}
V =  \frac{1}{(\lmax - 1)!} \sqrt{\sum_k \mu_{\pC_k \pC_k}^2 }  .
\end{equation}
 
The constant integral in \Eqref{eq:pregeom} can be interpreted as a measure of the 
volume of the manifold $K_\pC (s)$:
 \begin{equation} \label{eq:lln:geom}
 p \p{ \frac{S(\vec X | \pC )}{n} = s } = \Vol \p{K_\pC (s)}.
\end{equation}
Computing the volume of a general manifold can be quite difficult. However, $K_\pC (s)$
can be decomposed as an intersection of half-spaces and hyperplanes. It is thus a convex polytope,
a very specific subset of manifold which has been studied extensively. In particular, in order 
to compute the volume of a polytope a standard method consists in dividing the polytope into
a collection of simplices (i.e generalized triangles). For a given simplex $s$ with vertices $\{ v_1, \dots, v_k \}$ its volume can be computed by 
\begin{equation}
\Vol(s) = \det\p{v_2-v_1, \dots, v_k - v_1}.
\end{equation}
The volume of the whole polytope is then the sum of the volume of its decomposition in elementary simplices.
An interesting consequence of this is that the total volume of $K_\pC (s)$ depends only on the vertices of the polytope 
$K_\pC (s)$. As $K_\pC (s)$ is the intersection of $\Sp_{\pC }$ and the hyperplane $H_\pC (s)$, these vertices correspond 
to the intersection of the edges of $\Sp_{\pC }$ and the hyperplane $H_\pC (s)$. If we call $e_1,\dots,e_{D}$ the canonical
base of $\R^D$, the vertices of $\Sp_{\pC }$ are $e_{\pC_1},\dots, e_{\pC_{\lmax}}$. Then any $[e_{\pC_k}, e_{\pC_l}]$ segment is an edge of $\Sp_{\pC }$.
These segments are intersected by $H_\pC (s)$ if and only if their two end points lay on different sides of $H_\pC (s)$.
At a global level, if there are $k$ vertices on one side of $H_\pC (s)$ and $l$ on the other side, then $K_\pC (s)$ will
have $kl$ vertices. For instance, in dimension 4, the hyperplane $H_\pC (s)$ separates the $4$-simplex in either a $(1,4)$ 
configuration or a $(2,3)$ configuration. The first $(1,4)$ configuration corresponds to a tetrahedron with $4$ vertices.
The other $(2,3)$ configuration is a distorted triangular prism with $6$ vertices. In arbitrary dimension,
$K_\pC (s)$ is a kind of generalized prism\footnote{More precisely $K_\pC (s)$ is diffeomorph 
to the Cartesian product of a $|l-1|$-simplex and a $|r-1|$-simplex.}. 
The important result here is that the shape of $K_\pC (s)$ only changes when $H_\pC (s)$ crosses one of the $e_{\pC_k}$ vertices.
If we call $s^{\wedge}_t$ these crossing points, then on the intervals $(s^{\wedge}_t,s^{\wedge}_{t+1})$, the vertices $v^{k,l}$ of
$K_\pC (s)$ are affine functions of $s$ 
\begin{equation}
v^{k,l}(s) =  \frac{ s -  \mu_{\pC_l\pC_l} }{ \mu_{\pC_k\pC_k} - \mu_{\pC_l\pC_l} }  e_k + \frac{ s - \mu_{\pC_k\pC_k} }{ \mu_{\pC_l\pC_l} - \mu_{\pC_k\pC_k} }  e_l. 
\end{equation} 
Consequently, on the interval $(s^{\wedge}_t,s^{\wedge}_{t+1})$, $\Vol(K_\pC (s))$ is a polynomial function.
Hence, the limit distribution for the law of large numbers is a piecewise polynomial.
Moreover, it is possible to use symbolic computation to compute exactly the limit distribution from the means $\mu_{ii}$. For instance, if we arbitrarily choose 
\begin{equation}
\mu_{11} = \frac{35}{17}, \quad \mu_{22} =  1 \;,
\end{equation}
for our example, we have
\begin{equation}
 p \p{ \frac{S(\vec{X})}{N} = s } = 
\begin{cases}
 0 & s \in (-\infty,1) \;,\\
\frac{s}{3} - \frac{1}{3} & s \in [1,\frac{35}{17}) \;,\\
- \frac{18}{235} s + \frac{24}{47} & s \in [\frac {35}{17}, \frac{20} 3) \;,\\
 0 & s \in [ \frac{20} 3, +\infty ) \;.
\end{cases} 
\end{equation}

\begin{figure}
\centerline{ \includegraphics[width=0.7\columnwidth]{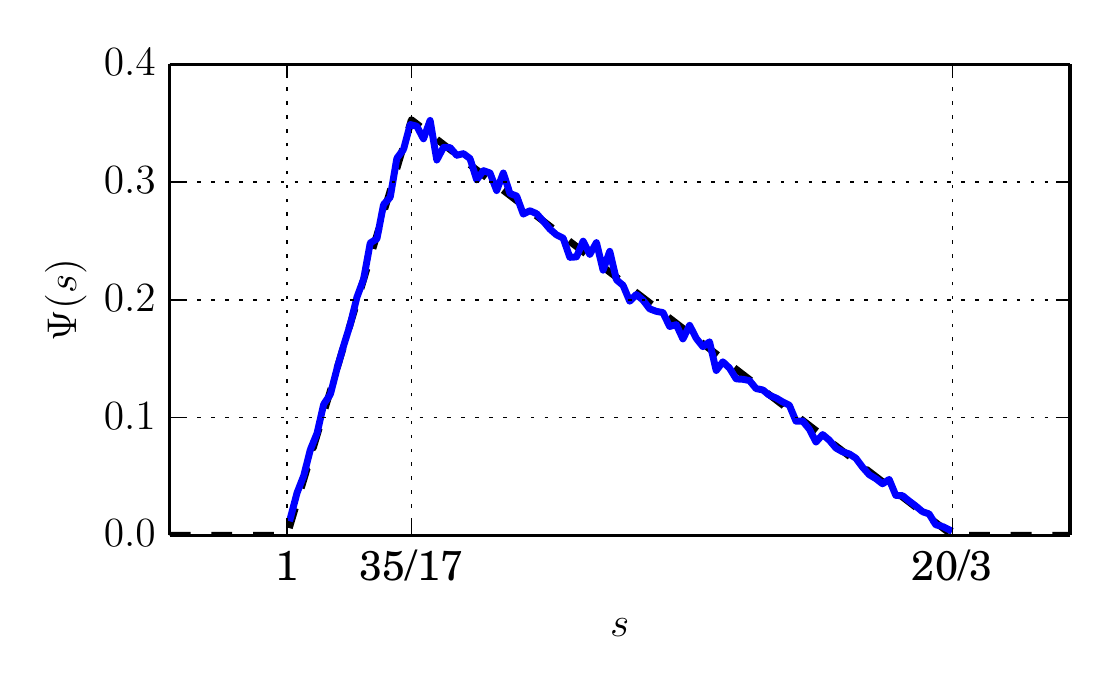} }
\caption{ \label{fig:lln} Limit law $\Psi(s)$ for sample average $s=S/N$ computed using a geometric method.
Dashed black line: theoretical distribution; Full blue line:
empirical histogram computed from $100000$ realizations of $\vec X$ with $N=1000$.
}
\end{figure}

\section{Conclusion}

In this contribution, we have shown that the sum of random variables with a matrix representation \Eqref{eq:Def}
generically converges to non-standard limit distributions, that we have characterized.
The existence of these non-standard limit distributions can be traced back to the presence of a form of ergodicity breaking of the underlying
hidden Markov chain. For any random variable with a matrix 
representation, it is possible to construct a reduced model
which encodes the non-ergodicity of the original model,
while preserving the limit distribution of the sum.
This mapping to a reduced model is a key element of our analytic results.
Through this approach, we have shown that the limit distribution of the sample mean 
can be determined as a discrete mixture of continuous mixtures of Dirac distributions. The standard law
of large numbers only holds if these mixtures reduce to a single Dirac distribution, which 
happens in particular if the hidden Markov chain is ergodic. Similarly, if the law of
large numbers holds, then the normalized centered sum converges to a discrete mixture of
continuous mixtures of Gaussian distributions. Since the resulting limit distributions have complicated expressions which are not straightforward to evaluate, we have proposed an algorithmic method to compute them.

The framework developed in this contribution can still be extended further. In particular,
we have restricted our study to the case of random variables with finite
variances. However, this restriction does not play any major role in our arguments. 
It should be possible to extend our results to the case of heavy-tailed distributions $\mP_{ij}(x)$. 
In a similar way, rather than studying the sum of random variables with a matrix representation,
we could have studied the extreme value statistics of such variables.
Indeed, the determination of the distribution of the maximum shares a significant number of formal properties with the sum \cite{Gyorgyi:2010b}. In particular, the maximum is, like the sum, a commutative, associative and $1$-homogeneous operator.
Exploiting these similarities, it should be possible to obtain similar results for the limit distributions
of the extrema of $\vec X$. Another major point of interest would be to extend these results to 
study the statistics of the number of particles in the stationary state of simple exclusion processes, that can be exactly described using a matrix product ansatz \cite{Evans07}. The difficulty here resides in the fact that the corresponding matrix product ansatz
differs slightly from the one we have proposed here (mostly due to the presence of negative coefficients in the matrix $\mE$) and cannot be modeled directly by a
Hidden Markov Model.


\appendix
\section{Periodic irreducible model} \label{app:periodic}
Periodic irreducible blocks $\mB{c}$ correspond to the case where
all the loops of the digraph $\gE(\mB{c})$ have a length which is a multiple of a base period
$P>1$:
\begin{equation}
(\mB{c}^{k})_{ii} > 0 \iff k \in  P \Z \;.
\end{equation}
Consequently, it is possible to partition the indices $\Xi_c$ in $P$ subsets $\sPer{o}$ with
$o \in  \Z/p\Z$ such that the edges of $\gE(\mB{c})$ only link indices from $\sPer{o}$ to $\sPer{o+1}$.
The chain $\sPth{c}$ cycles over the set $\sPer{o}$ with a period $P$ and therefore does not converge
to a stationary state. However, our aim is not to obtain a convergence result for the chain $\sPth{c}$ but for the transition frequencies
$\nu_{ij}$. The transition frequencies $\nu_{ij}$ are a global quantity that should not be influenced by the local periodic oscillation of $\sPth{c}$.
In particular, we can consider $\PGamma c o$, the $P$ subchains obtained by jumping over a period
\begin{equation}
\PGamma c o = \p{ \sPth{c}_{o}, \dots, \sPth{c}_{o+kP}, \dots, \sPth{c}_{o + P \floor{ (N t_c-o) /P } } } \;,
\end{equation}
where $\floor{n}$ denotes the integer part of $n$.
The chain $\PGamma c o$ corresponds to the hidden Markov chain of a matrix representation with structure matrix $\mB{c}^P$:
\begin{equation} \label{eq:prob:per}
P(\PGamma c o )  =
\frac{ \Lf\p{ 
	\mB{c}^{o-1} \left[ \prod_{k=1}^{\floor{(N t_c-o)/P}} (\mB{c}^P)_{{\PGamma c o}_{k} {\PGamma c o}_{k+1}} \right]
\mB{c}^{N t_c -  P \floor{(N t_c-o)/P} - o +1 }
} }
{ \Lf\p{ \mB{c}^{ N t_c } }}
 \;.
\end{equation}
Moreover,  if we call $\Pnu{c,o}$ the transition frequencies of the subchain $\PGamma c o$ then
\begin{equation} \label{eq:pnu:dec}
\bnu{c} = \frac 1 P \sum_{o=1}^{P} \Pnu{c,o}.
\end{equation}
The structure matrix $\mB{c}^P$ is no longer periodic. If we relabel the indices of $\mB{c}$ in order to make 
the $\sPer{o}$ contiguous, i.e.~to ensure that $\sPer{o} =  {s_o, s_0+1, \dots, d_o-1, d_o}$, then the matrix $\mB{c}^P$ reads
\begin{equation}
\mB{c}^P =
\begin{pmatrix}
D_{c,1} &        &   0   \\
    & \ddots &       \\
0   &        &   D_{c,P} \\  
\end{pmatrix}
\end{equation}
where $D_{c,o}$ are irreducible aperiodic square matrices of size $d_{c,o}$.
The block diagonal structure of $\mB{c}^P$ derives from the fact that after $P$ jumps, the periodic
chain $\sPth{c}$ goes back to its original set $\sPer{o}$. For two indices $o\ne o'$, there cannot be any transition between 
$\sPer{o}$ and $\sPer{o'}$ in the matrix $\mB{c}^P$.
In particular, if the final state of the subchain $\sPth{c}_{N t_c}=f$ belongs to the set $\sPer{\omega}$ 
then for a non-zero probability subchain $\sPth{c}$,
the chain $\PGamma c o$ stays inside the block $\sPer{\omega +o-N t_c}$:
\begin{equation} \label{eq:per:inv}
\forall  k,\quad {\PGamma c o}_k  \in  \sPer{\omega +o-N t_c} \;.
\end{equation}
Taking in account the property \Eqref{eq:per:inv}, 
\Eqref{eq:prob:per} simplifies to
\begin{equation}
P(\PGamma c o )  =
\frac{ \Lf\p{ 
	\mB{c}^{o-1} \left[ \prod_{k=1}^{\floor{(N t_c-o)/P}} (D_{c,\omega +o-N t_c})_{{\PGamma c o}_{k} {\PGamma c o}_{k+1}} \right]
\mB{c}^{N t_c -  P \floor{(N t_c-o)/P} - o +1 }
} }
{\Lf\p{ \mB{c}^{ N t_c } }}
\end{equation}

The subchain $\PGamma c o$ therefore converges to the stationary state $\vStat{c, \omega +o-N t_c}$ associated with the structure matrix 
$D_{c, \omega+o-N t_c}$. As in the aperiodic case, the transition frequencies are therefore
\begin{equation} \label{eq:pnu:conv}
\Pnu{c,o}_{ij} \as{\rightarrow}  \frac{ \fPEig{c, \omega +o-N t_c}_i \, \mB{c}_{ij} \, \vPEig{c , \omega +o-N t_c}_j }{\Lambda }
\end{equation}
where $\fPEig{c, o}$ and $\vPEig{c,o}$ are respectively the left- and right-eigenvectors of the block $D_{c,o}$ (embedded in the whole
vector space of $\mB{c}$).
Combining Eqs.~\eqref{eq:pnu:dec} and \eqref{eq:pnu:conv} yields
\begin{equation} \label{eq:pnu:conv:T}
 \bnu{c}   \as{\rightarrow} \frac 1 P \sum_{o=1}^{P}  \frac{  \fPEig{c,o}_i \mB{c}_{ij} \vPEig{c,o}_j }{\Lambda  } \;.
\end{equation}
The left and right eigenvectors  of $\mB{c}$ associated with $\Lambda$, respectively $\vEig{c}$ and $\fEig{c}$, are exactly
\begin{align}
\vEig{c}  =  \sum_{o=1}^P \vPtEig{c,o} \;,\\
\fEig{c}  =  \sum_{o=1}^P \fPtEig{c,o} \;.
\end{align}
Moreover, the support of the eigenvectors $\fPEig{c,o}$ and $\vPEig{c , o' }$ are disjoint if $o \ne o' $, 
consequently 
\begin{equation}
\fEig{c} \vEig{c}^T = \frac{1}{P} \sum_{o, o'} \fPtEig{c, o} \vPtEig{c, o'}^T =  \frac{1}{P} \sum_{o=1}^P \fPtEig{c,o}\vPtEig{c,o}^T .
\end{equation} 
Equation \eqref{eq:pnu:conv:T} therefore reads
\begin{equation} \label{eq:pnu:final}
 \bnu{c}_{i,j}  \as{\rightarrow}  \frac{  \fEig{c}_i \mB{c}_{ij} \vEig{c}_j} {\Lambda } \;.
\end{equation}
Hence the transition frequencies are exactly the same as the transition frequencies for the aperiodic case derived
in \Eqref{eq:nup:hom}.

\bibliographystyle{spmpsci}
\bibliography{SumStat,angeletti,LinAlg}

\end{document}

%% file: notations4.tex
\newcommand{\Eqref}[1]{Eq.~(\ref{#1})}
\newcommand{\Comment}[1]{}

\newcommand{\card}{\mathrm {card}}

\newcommand{\R}{\mathbb R}
\newcommand{\Z}{\mathbb Z}
\newcommand{\N}{\mathbb N}

\newcommand{\floor}[1]{\lfloor#1\rfloor}
\newcommand{\tr}{\mathrm{tr}}
\newcommand{\mId}{\mathrm I}

\newcommand{\len}[1]{|#1|}

\newcommand{\Sp}{ \mathcal M}
\newcommand{\Vol}{ \mathrm{Vol}}

\newcommand{\Esp}[1]{\left<#1\right>}

\newcommand{\conv}[1]{\underset{ #1 \rightarrow +\infty }{\rightarrow} }

\newcommand{\Prob}{P}
\newcommand{\iid}{\mathrm{i.i.d.}}
\newcommand{\as}[1]{\overset{\text{a.s}}{#1}}
	\newcommand{\Dist}[1]{\overset{\mathcal{D}}{#1}}
	\newcommand{\Distas}[1]{\overset{a.s.}{#1}}
	
	\newcommand{\convas}[1]{\Distas{\conv{#1}} }

	\newcommand{\lU}{\mathcal U}

\newcommand{\fracpow}[3]{\p{ \frac{#1}{#2}}^{#3}}
\newcommand{\p}[1]{\left(#1\right)} 
\newcommand{\repeated}[2]{ \underbrace{ #1,\dots,#1}_{ #2 \text{ times} } }
\newcommand{\fracLf}[2] { \frac{\Lf\p{#1}}{\Lf\p{#2}} }
\newcommand{\fLf}[1] { \fracLf{#1}{\mE^N} }

\newcommand{\mE}{\mathcal{E}}
\newcommand{\mP}{\mathcal{P}}
\newcommand{\mR}{\mathcal{R}}
\newcommand{\mA}{\mathcal{A}}
\newcommand{\Lf}{\mathcal{L}}
\newcommand{\mQ}[1]{M(#1)}

\newcommand{\ex}[1]{#1}
\newcommand{\mEx}{ \ex\mE}
\newcommand{\mAx}{ \ex\mA}

\newcommand{\gE}{\mathrm G}

\newcommand{\Conn}{\Pi}

	\renewcommand{\S}{S}
	\newcommand{\Imm}[1]{#1^{\circ}}
	
	\newcommand{\pnu}[1]{\nu_{#1}}

	\newcommand{\pC}{\mathcal C}
	\newcommand{\pfC}{\overline{\pC}}
	\newcommand{\pdC}{\hat{\pC}}
	\newcommand{\nc}{p}
	\newcommand{\sB}[1]{\Xi_{#1}}
	\newcommand{\sC}[1]{\Xi_{#1}}
	
	\newcommand{\mB}[1]{\mathcal B(#1)}
	\newcommand{\mbP}[1]{\mP^{|#1} }
	\newcommand{\imB}[1]{\Imm{\mathcal B}\p{#1}}

	\newcommand{\fEig}[1]{\rho(#1)}
	\newcommand{\vEig}[1]{\eta(#1)}

	\newcommand{\sPth}[1]{\Gamma^{|#1}}
	\newcommand{\vStat}[1]{ p_{\mathrm{st}}(#1) }
	\newcommand{\bnu}[1]{\nu(#1)}

	\newcommand{\cX}[1]{X^{[#1]}}

	\newcommand{\lmax}{{l_{\max}}}

	\newcommand{\NcUpB}{\p{\ln N}^2}
	\newcommand{\tcUpB}{ \frac{\NcUpB}{N}}
	\newcommand{\tcUpBl}{ \NcUpB/N }

	\newcommand{\redc}[2][]{#2^{\star#1}}
	\newcommand{\mEr}[1][]{\redc[#1]{\mE}}
	\newcommand{\mPr}[1][]{\redc[#1]{\mP}}
	\newcommand{\mAr}[1][]{\redc[#1]{\mA}}
	\newcommand{\Lfr}[1][]{\redc[#1]{\Lf}}
	\newcommand{\mRr}[1][]{\redc[#1]{\mR}}
	\newcommand{\pthr}[1][]{\redc[#1]{\Gamma}}
        \newcommand{\pCr}[1][]{\redc[#1]{\pC}}

	\newcommand{\mS}{\theta}
	\newcommand{\mSl}{\Theta}

	\newcommand{\PGamma}[2]{\sPth{#1,#2}}
	\newcommand{\sPer}[1]{\Theta_{#1}}
	\newcommand{\Pnu}[1]{\nu\p{#1}}

	\newcommand{\fPEig}[1]{\rho(#1)}
	\newcommand{\vPEig}[1]{\eta(#1)}

	\newcommand{\fPtEig}[1]{\rho(#1)}
	\newcommand{\vPtEig}[1]{\eta(#1)}